\documentclass[onecolumn,superscriptaddress,floatfix,nofootinbib,preprintnumbers
]{revtex4-1}
\usepackage{graphicx}
\usepackage{amsmath,amsfonts,mathrsfs,amssymb}
\usepackage{microtype}
\usepackage{bbm}
\usepackage[toc,page]{appendix}
\usepackage{hyperref}
\usepackage{subfigure}
\usepackage{braket}
\usepackage{stackengine}
\usepackage{array,float}
\usepackage{cancel}
\usepackage{color}
\usepackage{mathtools}
\usepackage{tikz}
\usepackage{indentfirst}
\usepackage{adjustbox}
\usepackage[many]{tcolorbox}
\usepackage{tikz-3dplot}
\usepackage{enumitem}
\usetikzlibrary{arrows,positioning} 

\tikzset{
    >=stealth',
    punkt/.style={rectangle, rounded corners, draw=black, very thick, text width=6.5em, minimum height=2em, text centered},
    pil/.style={->, thick, shorten <=2pt, shorten >=2pt,}
         }
\newcommand\irregularcircle[2]{
  \pgfextra {\pgfmathsetmacro\len{(#1)+rand*(#2)}}
  +(0:\len pt)
  \foreach \a in {10,20,...,350}{
    \pgfextra {\pgfmathsetmacro\len{(#1)+rand*(#2)}}
    -- +(\a:\len pt)
  } -- cycle
}
\begin{document}
\preprint{LA-UR-19-28837}
\preprint{YITP-19-84}
\preprint{CALT-TH-2019-009}

\begin{abstract}
We study the connections between three quantities that can be used as diagnostics for quantum chaos, i.e., the out-of-time-order correlator (OTOC), Loschmidt echo (LE), and complexity. We generalize the connection between OTOC and LE for infinite dimensions and extend it for higher-order OTOCs and multi-fold LEs. Novel applications of this intrinsic relation are proposed. We also investigated the relationship between a specific circuit complexity and LE by using the inverted oscillator model and made a conjecture about their relationship. 
These relationships signal a deeper connection between these three probes of quantum chaos. 
\end{abstract}

\title{Towards the Web of Quantum Chaos Diagnostics}
\vskip 1in

\author{Arpan Bhattacharyya}
\email{abhattacharyya@iitgn.ac.in}
\affiliation{Indian Institute of Technology,Gandhinagar,Gujarat 382355, India}
\affiliation{Center for Gravitational Physics,
Yukawa Institute for Theoretical Physics (YITP), Kyoto University, Kitashirakawa Oiwakecho, Sakyo-ku, Kyoto 606-8502, Japan.}

\author{Wissam Chemissany}
\email{wissamch@caltech.edu}
\affiliation{Institute for Quantum Information and Matter, California Institute of Technology,\\
1200 E California Blvd, Pasadena, CA 91125, USA.}

\author{S. Shajidul Haque}
\email{shajidhaque@gmail.com}
\affiliation{Department of Mathematics and Applied Mathematics, University of Cape Town,
Private Bag, Rondebosch, 7701, South Africa.}

\author{Bin Yan}
\email{byan@lanl.gov}
\affiliation{Center for Nonlinear Studies, Los Alamos National Laboratory, Los Alamos, NM 87544, USA}
\affiliation{Theoretical Division, Los Alamos National Laboratory, Los Alamos, NM 87544, USA}

\vskip.5mm

\maketitle
\tableofcontents

\section{Introduction}


Characterizing the nature of quantum chaos~\cite{Haake2006-ic} in quantum many body systems can be challenging. This area of research is versatile and appears in many branches of theoretical and experimental physics. This has spurred a renewed interest in the quest for a quantum version of a classical chaos in the last few years. For recent developments, interested readers are referred to~\cite{HunterJones2018ChaosAR} and the references therein. Quantum chaos has found applications and received considerable attention across physical disciplines such as condensed matter physics, quantum information theory and high energy physics, in particular, in the context of black hole and holography~\cite{Jahnke:2018off}. Several diagnostic tools have been proposed to quantify it's diverse applications.

Over time, the endeavors to improve the current diagnostic gadgets and develop new ones have gone a long way. The out-of-time-order correlator (OTOC) \cite{larkin1969quasiclassical,kitaevProc2015} has been intensively utilized to examine chaotic behaviour 
providing deeper understanding for long-standing problems. Loschmidt echo (LE), introduced as another powerful toolkit \cite{goussev2012loschmidt,gorin2006dynamics}, has also played a pivotal role in demystifying the structure of (quantum) chaos. Very recently, a quantum information theoretic tool  called quantum circuit complexity has joined the club of quantum chaos diagnostics~\cite{Susskind:2018tei,Magan:2018nmu,Balasubramanian:2019wgd,Yang:2019iav}. In \cite{Ali2019-ks}, the authors have shown that quantum complexity for a specific type of quantum circuit, namely, circuit complexity~\cite{NL1,Jefferson:2017sdb,Chapman, Khan, Hackl:2018ptj}, can capture the chaotic features.

There are serious indications that this proposed chaos \emph{quantifiers} are related to each other. For instance, there had been a strong belief that the OTOC and LE are connected to each other due to the intrinsic nature of the echo of the OTOC~\cite{Chenu2018-sm,Chenu2019-ue,kurchan2018quantum}. Indeed, in~\cite{zurek} a major step has been taken to establish a direct link. It is worth mentioning that, previously there had been several attempts~\cite{romero2019regularization, kurchan2018quantum} to achieve the same goal, but all of them are resorted to some variants of the OTOC or specific choice of operators. In~\cite{Ali2019-ks} a close connection between certain OTOC and complexity has been proposed. These results indicate a deeper connection between these diagnostics and provide motivations to explore it further. 

In this paper we would like to initiate a program that ultimately intends to investigate towards a complete web of quantum chaos diagnostics. To be specific, the aim of this paper is two-fold; i) first, generalize the OTOC-LE connection of~\cite{zurek} for infinite dimensional system and extend it to $k$ multi-fold and provide examples, ii) second, explore the relationship between LE and complexity. 
To investigate the LE-complexity relation we use the inverted harmonic oscillator model for computing complexity and establish its connection to a particular type of LE. Finally we will comment and speculate on the possible ways to go beyond this example and highlight some future directions. 



\section{Loschmidt echo and OTOC}
We start with an introduction of Loschmidt echo and the regular $4$-point OTOC. Then we will discuss the general properties of these two quantities, as well as the intrinsic connection between them. Our first result is to generalize the link between the regular OTOC and LE to higher-order OTOCs and a echo quantity with multiple loops. This leads to a wide range of novel applications.

\begin{figure}[ht]\label{fig_LE}
\begin{tikzpicture}[node distance=1cm, auto]
\node (A) at (0, 0) {$\ket{\psi_0}$};
\node (B) at (4, 0) {$\ket{\psi(t)}$};

\node[] at (0,-0.5) {0};
\node[] at (4,-0.5) {$t$};
\node[] at (2,0.5) {$e^{-i\mathcal{H}_1t}$};
\node[] at (2,-0.5) {$e^{i\mathcal{H}_2t}$};

\draw[->, thick] (0.5,0.5) .. controls (2,1) .. (3.5,0.5);
\draw[->, thick] (3.5,-0.5) .. controls (2,-1) .. (0.5,-0.5);
\end{tikzpicture}
\caption{LE as an echo quantity measures how much a quantum state is recovered by an imperfect time reversal.}
\end{figure}
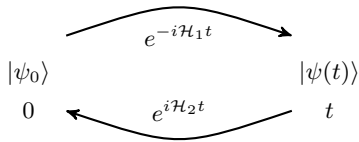
\vspace{2mm}
\noindent
The LE is formally defined as \cite{gorin2006dynamics}
\begin{equation}
    M(t)=|\bra{\psi_0}e^{i\mathcal{H}_2t}e^{-i\mathcal{H}_1t}\ket{\psi_0}|^2,
\end{equation}
where $\ket{\psi_0}$ is the initial state of a quantum system, $\mathcal{H}_1$ and $\mathcal{H}_2$ are two slightly different Hamiltonian, e.g., $\mathcal{H}_1=\mathcal{H}_0$ is the unperturbed Hamiltonian, and $\mathcal{H}_2=\mathcal{H}_0+V$ with $V$ a small perturbation. 

One can interpret the LE in two ways. First, it can be considered as an ``echo'' process. It quantifies how much of the complex system is recovered upon applying an imperfect time-reversal, as sketched in Fig.~\ref{fig_LE}. The other way is to interpret it as the overlap (the ``distance'') between two wavefunctions (``trajectories'') evolving under slightly different dynamics. This is analogous to the classical notion of chaos, though in the latter case perturbations are applied to the initial condition in the classical phase space, while in the quantum case the perturbations are applied to the Hamiltonian. (Due to the fundamental unitary dynamics in quantum systems, any small perturbations on the initial wavefunctions remain unchanged during time evolution.) In this sense, the LE is related to the \emph{butterfly effect}, so one can consider it as a diagnostic for chaos.\\


The regular 4-point OTOC is formally defined as
\begin{equation}
    F_\beta(t)=\langle W^{\dagger}(t)V^{\dagger}(0)W(t)V(0) \rangle_\beta,
\end{equation}
Here the average is taken over a thermal state at inverse temperature $\beta$. $W$ and $V$ are two local operators on distinct local subsystems. $W(t)\equiv e^{-iHt}We^{iHt}$ is the Heisenberg evolution of the operator $W$. The OTOC has been extensively studied in various context and different variants of it has been proposed \cite{2016arXiv160801914F,Cotler:2017jue}. 
We note the following universal features of the OTOC:
\begin{itemize}
    \item When $W$ and $V$ are both Hermitian and unitary, the OTOC is related to the squared commutator
    \begin{equation}
        F_\beta(t)=1-\frac{1}{2}\langle[W(t),V]^2\rangle.
    \end{equation}
    Two local operator $W$ and $V$ commute at $t=0$. The Heisenberg evolution converts $W(t)$ into a global operator; the commutator hence fails to vanish and induces decay of the OTOC. For chaotic dynamics the OTOCs exhibits fast decays.
    \item The OTOC has several decay regimes. At early stage before the Ehrenfest time scale (also known as the scrambling regime \cite{Maldacena:2015waa}), the decay of OTOC is manifested as an exponential growth, $1-\delta e^{\lambda t}$, where $\delta \ll 1$. This type of decay certainly does not converge, and will switch to a pure exponential decay (intermediate regime) before saturation. In the asymptotic regime (late time) the OTOC typically shows model-dependent power law behaviors. In the scrambling regime, the exponential growth rate is conjectured to be bounded by the temperature, i.e., $\lambda\le 2\pi/\beta$ from holography \cite{Maldacena:2015waa}. 
    \item The OTOC recovers the essential pieces of the classical notion of chaos in phase space. A heuristic way to see this is to look at the semi-classical limit \cite{Maldacena:2015waa} for the choice of operators $W(t)=q(t)$ and $V=p$, where $q$ and $p$ are conjugate pair of variables. In the semi-classical limit, the commutator reduces to the Poisson bracket. This gives $[q(t),p]\rightarrow i\hbar\{q(t),p\}\sim i\hbar \partial q(t)/\partial q(0)$, which grows as $e^{\lambda_L t}$ with $\lambda_L$ the Lyapunov exponent. For various systems with classical counterparts, e.g., the kicked rotor \cite{Rozenbaum2017-fh} or the cat map \cite{Garcia-Mata2018-xe}, the decay rates of the OTOC were shown to match the classical Lyapunov exponents. 
    \item Different choices of operators of $W$ and $V$ share common features of their OTOCs. For complex enough systems, the OTOC is expected to be not sensitive to the particular form of the operators, especially when we are interested in extracting the universal characteristics. This makes it possible to get rid of the operator dependence by averaging over all operators of given subsystems. In the following sections we explore the consequences of this averaging procedure. 
\end{itemize}

\begin{figure}[H]
\centering
\begin{tikzpicture}
\coordinate (c) at (5.5,0);
\coordinate (d) at (8.5,0);
\draw[blue, thick, rounded corners=1mm] (c) \irregularcircle{0.4cm}{0.6mm};
\draw [thick,  fill= gray](5.5,-0) circle(0.1cm);
\node[] at (5.5, -1.8){$W$};
\draw [thick, fill= gray](7.75,0.5) circle(0.1cm);
\draw [thick, fill= gray](7.9,0) circle(0.1cm);
\draw [thick, fill= gray](7.8,-0.6) circle(0.1cm);
\draw [thick, fill= gray](8.2,0.3) circle(0.1cm);
\draw [thick, fill= gray](8.5,-0.5) circle(0.1cm);
\draw [thick, fill= gray](9.25,0.5) circle(0.1cm);
\draw [thick, fill= gray](9.5,0) circle(0.1cm);
\draw [thick, fill= gray](9.25,-0.2) circle(0.1cm);
\draw [thick, fill= gray](8.75,0.425) circle(0.1cm);
\node[] at (8.5, -1.8){$V$};
\draw[cyan,  thick, rounded corners=1mm] (d) \irregularcircle{1.3cm}{1mm};
\end{tikzpicture}
\caption{Local structure of the total system and the choice of subsystems in the 4-point OTOC.}\label{fig_AB}
\end{figure}
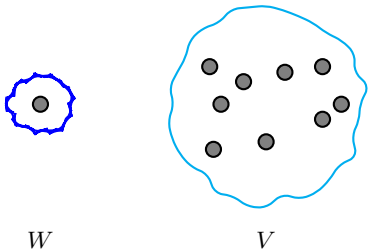

\vspace{2mm}

As noted in the previous section, the insensitivity to the choice of operators allows one to extract the universal features of OTOC by taking the average over a given set of operators. The averaging procedure has been considered for different variants of the OTOCs. When restricted to the original form with local operators, there exists a strong relation between the Loschmidt echo and the 4-point OTOC: Without losing the local structure of a many-body system, the supports of two operators $W$ and $V$ are chosen as two distinct subsystems $\mathcal{A}$ and $\mathcal{B}$, where $\mathcal{A}$ is a small subsystem, while $\mathcal{B}$ is the complement of $\mathcal{A}$ to the total system, as illustrated in Fig. \ref{fig_AB}. We then take the average of the two operators over the set of all unitaries  on the two fixed subsystems with the ``largest randomness'', i.e., with respect to the Haar measure.  
It has been demonstrated in \cite{zurek} that the OTOC and LE are ultimately related as
\begin{equation}
         \int\limits_{Haar} dWdV\langle W^{\dagger}(t)V^{\dagger}W(t)V\rangle_{\beta=0}
    \approx \bigg|\langle e^{i\mathcal{H}_\mathcal{B}} \times e^{-i(\mathcal{H}_\mathcal{B}+P)t}\rangle_{\beta=0}\bigg|^2.
\end{equation}
Here the Hamiltonian of the larger subsystem $\mathcal{B}$ plays the role of the unperturbed Hamiltonian; and the perturbation $P$ naturally emerges from the interaction between the two subsystems. It is the projection of the interaction to the Hilbert space of the subsystem $\mathcal{B}$ (See Ref.\cite{zurek} or the following derivations for the construction of the effective perturbations). This relations was shown to be valid in both the scrambling and the intermediate decay regime. In the following section, we will further generalize this result for 2k-point OTOC to the 2(k-1)-fold Loschmidt echo, and also to the case where both of the operators $W$ and $V$ are supported on small local subsystems. For simplicity, we will restrict ourselves to the case of infinite temperatures. With proper regularization of the thermal state, this relation generalizes to finite temperature as well, using techniques developed in Ref.~\cite{zurek}.
\subsection{2k-OTOC and 2(k-1)-fold echo}
We will start with a formal definition of the $2k$-point OTOC. Then we will demonstrate that it is linked to a LE with $2(k-1)$ forward and backward loops.

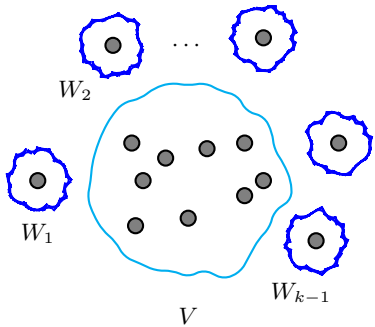
\begin{figure}[H]
\centering
\begin{tikzpicture}
\coordinate (c) at (6.5,0);
\coordinate (d) at (8.5,0);
\coordinate (c1) at (7.5,1.8);
\coordinate (c2) at (9.5,1.9);
\coordinate (c3) at (10.5,0.5);
\coordinate (c4) at (10.2,-0.8);
\draw[blue, thick, rounded corners=1mm] (c) \irregularcircle{0.4cm}{0.6mm};
\draw[blue, thick, rounded corners=1mm] (c1) \irregularcircle{0.4cm}{0.6mm};
\draw[blue, thick, rounded corners=1mm] (c2) \irregularcircle{0.4cm}{0.6mm};
\draw[blue, thick, rounded corners=1mm] (c3) \irregularcircle{0.4cm}{0.6mm};
\draw[blue, thick, rounded corners=1mm] (c4) \irregularcircle{0.4cm}{0.6mm};
\draw [thick,  fill= gray](6.5,-0) circle(0.1cm);
\draw [thick,  fill= gray](7.5,1.8) circle(0.1cm);
\draw [thick,  fill= gray](9.5,1.9) circle(0.1cm);
\draw [thick,  fill= gray](10.5,0.5) circle(0.1cm);
\draw [thick,  fill= gray](10.2,-0.8) circle(0.1cm);
\node[] at (6.5, -0.7){$W_1$};
\node[] at (7, 1.2){$W_2$};
\node[] at (8.5, 1.8){$\cdots$};
\node[] at (10, -1.5){$W_{k-1}$};
\draw [thick, fill= gray](7.75,0.5) circle(0.1cm);
\draw [thick, fill= gray](7.9,0) circle(0.1cm);
\draw [thick, fill= gray](7.8,-0.6) circle(0.1cm);
\draw [thick, fill= gray](8.2,0.3) circle(0.1cm);
\draw [thick, fill= gray](8.5,-0.5) circle(0.1cm);
\draw [thick, fill= gray](9.25,0.5) circle(0.1cm);
\draw [thick, fill= gray](9.5,0) circle(0.1cm);
\draw [thick, fill= gray](9.25,-0.2) circle(0.1cm);
\draw [thick, fill= gray](8.75,0.425) circle(0.1cm);
\node[] at (8.5, -1.8){$V$};
\draw[cyan,  thick, rounded corners=1mm] (d) \irregularcircle{1.3cm}{1mm};
\end{tikzpicture}
\caption{Local structure of the total system and the choice of subsystems in the 2k-OTOC.}\label{fig_2kOTOC}
\end{figure}

The regular 4-point OTOC, 
\begin{equation*}
    \langle W^{\dagger}(t)V^{\dagger}(0)W(t)V(0) \rangle,
\end{equation*} 
probes the spreading of the local operator $W$ over the entire system. The work of Refs. \cite{Roberts2017-tt, Shenker2014-ep} suggests the study of the generalized 2k-OTOC, defined as
\begin{equation}\label{2kotoc}
\begin{aligned}
         &\langle\mathcal{W}^{\dagger} V^{\dagger}(0)\mathcal{W} V(0)\rangle \\
         =& \langle W_1^{\dagger}(t_1)...W_{k-1}^{\dagger}(t_{k-1})
     V^{\dagger}(0)W_{k-1}(t_{k-1})... W_1(t_1)V(0)\rangle,
\end{aligned}
\end{equation}
where $\mathcal{W} \equiv W_{k-1}(t_{k-1})... W_1(t_1)$ indicates the ordering of the operators in the correlator. 
Note that there are other 
definitions of the 2k-OTOC, such as the ones used in \cite{Roberts2017-tt} to probe k-designs and the one connected to the spectral form factors in \cite{Cotler2017-oi}. The operators in the correlator could be interpreted as either global as in \cite{Roberts2017-tt,Cotler2017-oi}) or local operators as in \cite{Shenker2014-ep}) for different purposes.

We first focus on the most restricted choice, i.e., the operator $W_k$'s are all local operators applying on distinct local (and small) subsystems, such that the 2k-OTOC mentioned in (\ref{2kotoc}) probes the scrambling of multiple local perturbations. We choose $V$ as an operator on the complement of the $k-1$ local subsystems. The structure of this 2k-OTOC is illustrated in Fig.~\ref{fig_2kOTOC}. We denote
\begin{equation*}
    W_k(t_k) \equiv U_k^{\dagger} W_k U_k \equiv \tilde{W}_k,
\end{equation*}
where $U_k=e^{i H t_k}$ is the evolution operator. Consider the averaged 2k-OTOC with respect to the Haar integral at infinite temperature,
\begin{equation}
\begin{aligned}
     &\int_{Haar} dW_1... dW_{k-1}dV\ \langle \tilde{W}_1^{\dagger}... \tilde{W}_{k-1}^{\dagger} V^{\dagger}
        \tilde{W}_{k-1}... \tilde{W}_1V\rangle_{\beta=0}\\
    =&\frac{1}{d}\, {\rm Tr} \int_{Haar} dW_1... dW_{k-1}dV (\tilde{W}_1^{\dagger}... \tilde{W}_{k-1}^{\dagger} V^{\dagger}
     \tilde{W}_{k-1}... \tilde{W}_1V).
        \end{aligned}
\end{equation}
Here $d$ is the dimension of the total Hilbert space. The particular ordering of operators in the integrand allows us to perform the integral one-by-one, e.g., the inner-most integral for the $W_{k-1}$ operators can be computed first (see Appendix A for Haar average over subsystems):
\begin{widetext}
\begin{equation}
\begin{aligned}
 &\int dW_{k-1}\  \bigg(\tilde{W}_1^{\dagger}... \tilde{W}_{k-1}^{\dagger} V^{\dagger} \tilde{W}_{k-1}... \tilde{W}_1V\bigg) \\
= & \int dW_{k-1}\quad  \bigg(\tilde{W}_1^{\dagger} ... \tilde{W}_{k-2}^{\dagger} U_{k-1}^{\dagger} W^{\dagger}_{k-1}  U_{k-1} V^{\dagger} U^{\dagger}_{k-1} W_{k-1} U_{k-1} \tilde{W}_{k-2}... \tilde{W}_1 V \bigg)\\
= &\ \frac{1}{d_{k-1}} \tilde{W}_1^{\dagger} ... \tilde{W}_{k-2}^{\dagger} U_{k-1}^{\dagger} {\rm Tr}_{k-1} \bigg( U_{k-1} V^{\dagger} U^{\dagger}_{k-1}\bigg) U_{k-1} \tilde{W}_{k-2}... \tilde{W}_1 V,
\end{aligned}
\end{equation}
\end{widetext}
where $Tr_{k-1}$ represents the partial trace over the subsystem $k-1$, and $d_k$ is the dimension of the local Hilbert space supporting operator $W_k$. Performing the integral for all the $W$ operators gives
\begin{widetext}
\begin{align}
\begin{split}
        &\frac{1}{d}\,{\rm Tr} \int\limits_{Haar} dW_1... dW_{k-1}dV\bigg(\tilde{W}_1^{\dagger}...\tilde{W}_{k-1}^{\dagger} V^{\dagger} \tilde{W}_{k-1}... \tilde{W}_1V\bigg) \\
        = & \frac{1}{d}\frac{1}{d_{1}... d_{k-1}}\, {\rm Tr}\ \int \ dV\, U_1^{\dagger}\,{\rm Tr}_1 \bigg( U_1 ... U_{k-2}^{\dagger}
          {\rm Tr}_{k-2}\bigg(U_{k-2} U_{k-1}^{\dagger}{\rm Tr}_{k-1}\bigg(U_{k-1}V^{\dagger} U_{k-1}^{\dagger} \bigg)U_{k-1} U_{k-2}^{\dagger} \bigg) U_{k-2} ... U_1^{\dagger} \bigg)U_1 V\\
        = & \frac{1}{d}\frac{1}{d_{1}... d_{k-1}}\, {\rm Tr}\int dV\,  {\rm Tr}_1 \bigg\{ U_1 ... U_{k-2}^{\dagger}{\rm Tr}_{k-2}\bigg[U_{k-2} U_{k-1}^{\dagger}
        {\rm Tr}_{k-1}\bigg(U_{k-1}V^{\dagger} U_{k-1}^{\dagger} \bigg)U_{k-1} U_{k-2}^{\dagger} \bigg]U_{k-2} ... U_1^{\dagger} \bigg\}\, U_1 V U_1^{\dagger}
\end{split}
\end{align}
\end{widetext}
Let us define the following:
\begin{widetext}
\begin{align}
    \begin{split}
        {\rm Tr}_1 \bigg\{ U_1 ... U_{k-2}^{\dagger}{\rm Tr}_{k-2}\bigg[U_{k-2} U_{k-1}^{\dagger}
        {\rm Tr}_{k-1}\bigg(U_{k-1}V^{\dagger} U_{k-1}^{\dagger} \bigg)U_{k-1} U_{k-2}^{\dagger} \bigg]U_{k-2} ... U_1^{\dagger} \bigg\}&\equiv A,\\
        U_1 V U_1^{\dagger}&\equiv B.
    \end{split} \nonumber
\end{align}
\end{widetext}
Using the same trick provided for the 4-point OTOC in \cite{zurek}, the partial traces in piece-$A$ can be evaluated one-by-one (see Appendix A). For instance, the inner-most partial trace is
\begin{equation*}
        {\rm Tr}_{k-1}\left(U_{k-1}V^{\dagger} U_{k-1}^{\dagger} \right)={\rm Tr}_{k-1}\left(e^{-iHt_{k-1}}V^{\dagger} e^{iHt_{k-1}}\right) \approx d_{k-1}\times\frac{1}{N_{k-1}}\sum_{P_{k-1}} e^{-i(H_V+P_{k-1})t_{k-1}}V^{\dagger} e^{i(H_V+P_{k-1})t_{k-1}}.
\end{equation*}
Here $N_{k-1}$ is the number of different $P_{k-1}$ operators, which serve as the perturbations. The summation range over all of them. These noisy operators emerge from the interaction between the $k-1$'s subsystem with the rest of the total system (See Appendix A for details).
%
Note that the LHS of the above equation, after tracing over the $(k-1)$'th subsystem, is an operator that involves not only the subsystem-V, but also subsystems $1,2,...,k-2$. However, we assume that it only evolves (under noises) in subsystem-V and it does not ``leak'' to other subsystems. 
%
\begin{equation*}
    U_{k-2}U^{\dagger}_{k-1}\equiv e^{-iHt_{k-2}}e^{iHt_{k-1}}=e^{-iH(t_{k-2}-t_{k-1})},
\end{equation*}
the above procedure for partial tracing can be repeated to all partial traces, which give the following expression for $A.$
Before proceeding further, let us define the following,
\begin{align}
    e^{-i(H_V+P_1)(t_1-t_2)}\cdots e^{-i(H_V+P_{k-2})(t_{k-2}-t_{k-1})}e^{-i(H_V+P_{k-1})t_{k-1}}\equiv D. \nonumber
\end{align}
Then we get, 
\begin{equation}
         A=d_1\dots d_{k-1}\frac{1}{N_1\dots N_{k-1}}\sum_{P_1,...,P_{k-1}}\,D\, V^{\dagger} \, D^{\dagger}.
\end{equation} 
Finally, 
\begin{equation}
      \  \frac{1}{d}\frac{1}{d_{1}... d_{k-1}} {\rm Tr}\ \int \ dV\  A\,B
      = \frac{1}{d}\frac{1}{N_{1}... N_{k-1}} {\rm Tr}\ \int \ dV\ \sum_{P_1,...,P_{k-1}} D V^{\dagger} D^{\dagger} U_1 V U^{\dagger}_1. \label{new1}
\end{equation} 
As has been discussed before, $U_1VU^{\dagger}_1$ is a global operator, while $DV^{\dagger} D^{\dagger}$ is an operator with support on system-$V$ only. Thus the trace in the above equation can be evaluated with two partial traces ${\rm Tr}={\rm Tr}_V {\rm Tr}_{\bar{V}}$, namely, ${\rm Tr}[(M_V\otimes \mathbb{I}_{\bar{V}})N_{V\bar{V}}]={\rm Tr}_V[M_V {\rm Tr}_{\bar{V}}(N_{V\bar{V}})]$. Denote $d_{1}... d_{k-1}\equiv d_{\bar{V}}$, which is the dimension of the Hilbert space of the subsystem complementary to subsystem-$V$. The right hand side of the  equation (\ref{new1}) continues as
\begin{equation}
\begin{aligned}   
      =& \frac{1}{d}\frac{1}{N_{1}...N_{k-1}} \sum_{P_1,...,P_{k-1}} {\rm Tr}_V \bigg[\int dV D V^{\dagger} D^{\dagger} {\rm Tr}_{\bar{V}}\bigg(U_1 V U^{\dagger}_1\bigg)\bigg],\\
      =&\frac{1}{N_1...N_{k-1}}\frac{1}{d_V}\frac{1}{d_{\bar{V}}} \sum_{P_1,...,P_{k-1}} {\rm Tr}_V \bigg[\int dV D V^{\dagger} D^{\dagger}
     \bigg(\frac{1}{N_0}\sum _{P_0}e^{-i(H_V+P_0)t_1} Ve^{i(H_V+P_0)t_1}\bigg)\bigg],\\
      =& \frac{1}{N_0...N_{k-1}} \frac{1}{d_V} \sum_{P_0,...,P_{k-1}} {\rm Tr}_V \bigg(\ \int \ dV\  D V^{\dagger} D^{\dagger} e^{-i(H_V+P_0)t_1} Ve^{i(H_V+P_0)t_1}\bigg),\\
      =&\frac{1}{N_0...N_{k-1}} \frac{1}{d^2_V} \sum_{P_0,...,P_{k-1}} \bigg|{\rm Tr}(e^{i(H_V+P_0)t_1} D)\bigg|^2,\\
      =& \frac{1}{N_0...N_{k-1}} \frac{1}{d^2_V} \sum_{P_0,...,P_{k-1}}\Bigg|{\rm Tr} \bigg[e^{i(H_V+P_0)t_1} e^{-i(H_V+P_1)(t_1-t_2)}... e^{-i(H_V+P_{k-2})(t_{k-2}-t_{k-1})} e^{-i(H_V+P_{k-1})t_{k-1}}\bigg]\Bigg|^2.
\end{aligned}\label{new2}
\end{equation}
$P_1,\dots,P_{k-1}$ are perturbations emerge from the tracing our the subsystems -$1,\dots,k-1$; and they have, respectively. $P_0$ emerges from tracing out the subsystem $\bar{V}$. For complex systems, the structure of these perturbation operators are not essential. Hence, we can eliminate the average over all the perturbations and treat each $P_i$ as a constant perturbation instead of a variable.  Finally we get,
\begin{equation}
\begin{aligned}
    \frac{1}{d} {\rm Tr} \int\limits_{Haar} dW_1... &dW_{k-1}dV\bigg(\tilde{W}_1^{\dagger}...\tilde{W}_{k-1}^{\dagger} V^{\dagger} \tilde{W}_{k-1}... \tilde{W}_1V\bigg)= \frac{1}{d_V^2}\Bigg|{\rm Tr} \bigg[e^{i(H_V+P_0)t_1}D\bigg]\Bigg|^2,\\&
     = \Bigg|\langle\bigg[e^{i(H_V+P_0)t_1}e^{-i(H_V+P_1)t_1}e^{i(H_V+P_1)t_2}\cdots e^{-i(H_V+P_{k-2})t_{k-2}}e^{i(H_V+P_{k-2})t_{k-1}} e^{-i(H_V+P_{k-1})t_{k-1}}\bigg]\rangle_{\beta=0}\Bigg|^2.
\end{aligned}
\end{equation}
This is the expected Loschmidt echo with $2(k-1)$ loops.

\subsection{Local OTOCs}\label{sec:localOTOC}

\begin{figure}[H]
\centering
\begin{tikzpicture}
\coordinate (c) at (6,0);
\coordinate (c1) at (11,0);
\coordinate (d) at (8.5,0);
\draw[blue, thick, rounded corners=1mm] (c) \irregularcircle{0.4cm}{0.6mm};
\draw[blue, thick, rounded corners=1mm] (c1) \irregularcircle{0.4cm}{0.6mm};
\draw [thick,  fill= gray](6,-0) circle(0.1cm);
\draw [thick,  fill= gray](11,-0) circle(0.1cm);
\node[] at (6, -1.8){$W$};
\draw [thick, fill= gray](7.75,0.5) circle(0.1cm);
\draw [thick, fill= gray](7.9,0) circle(0.1cm);
\draw [thick, fill= gray](7.8,-0.6) circle(0.1cm);
\draw [thick, fill= gray](8.2,0.3) circle(0.1cm);
\draw [thick, fill= gray](8.5,-0.5) circle(0.1cm);
\draw [thick, fill= gray](9.25,0.5) circle(0.1cm);
\draw [thick, fill= gray](9.5,0) circle(0.1cm);
\draw [thick, fill= gray](9.25,-0.2) circle(0.1cm);
\draw [thick, fill= gray](8.75,0.425) circle(0.1cm);
\node[] at (11, -1.8){$V$};
\draw[cyan,  thick, rounded corners=1mm] (d) \irregularcircle{1.3cm}{1mm};
\end{tikzpicture}
\caption{Local structure of the total system and the choice of subsystems in the local 4-point OTOC.}\label{fig_localOTOC}
\end{figure}
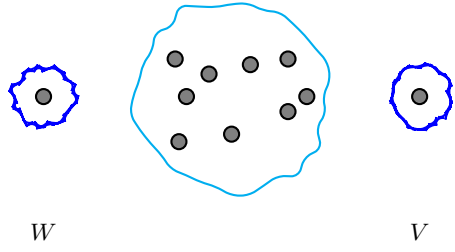

In the previous section, we have established the OTOC-LE correspondence for one type of generalized 2k point OTOC. This 2k-OTOC involved many local operators, and a giant operator over the compliment system of the other local operators (see Fig.~\ref{fig_2kOTOC}) for illustration. In this section, we consider a even more general case, where the OTOC only involves operators on small local subsystems. For simplicity, we only consider the case of the 4-point OTOC and infinite temperature. Generalization to the multi-point case follows from the techniques developed in the previous section. The four point OTOC at infinite temperature is 
\begin{equation}
    \langle W^{\dagger}(t)V^{\dagger}(0)W(t)V(0) \rangle_{\beta=0} = \frac{1}{d} {\rm Tr} \left[ W^{\dagger}(t)V^{\dagger}(0)W(t)V(0)\right],
\end{equation}
with the operators $W$ and $V$ constrained to local subsystems (see Fig.~\ref{fig_localOTOC} for illustration.).

Averaging $W$ and $V$ over all unitaries on the corresponding subsystems, we get
\begin{equation}
\begin{aligned}
\overline{\text{OTOC}}_{4}&=\int_{Haar}dV\,dW\,\langle W^{\dagger}(t)V^{\dagger}(0)W(t)V(0) \rangle\\
    &=\frac{1}{d}{\rm Tr}\int dV\,dW\, e^{iHt}  W^\dag e^{-iHt}V^\dag e^{iHt}We^{-iHt}V\\
    &=\frac{1}{d\cdot d_V} {\rm Tr} \int dW\, |{\rm Tr}_V e^{iHt}  We^{-iHt}|^2.
\end{aligned}
\end{equation}
In the above equation, the reduced evolution of operator $W$ (partial trace ${\rm Tr}_V$ over the subsystem of $V$) can be approximated with
\begin{equation}
    {\rm Tr}_V e^{iHt}  We^{-iHt} = d_V  e^{i(H_{\bar{V}}+P_V)t}  W e^{-i(H_{\bar{V}}+P_V)t},
\end{equation}
where $d_V$ is the dimension of the support of operator $V$. $H_{\bar{V}}$ is the Hamiltonian of the subsystem excluding the support of $V$. Hence,
the evaluation of the four point OTOC continues as
\begin{equation}
\begin{aligned}
\overline{\text{OTOC}}_{4}
=&\frac{d_V}{d} {\rm Tr} \int dW \  W^\dag e^{-i(H_{\bar{V}}+P_V)t}e^{i(H_{\bar{V}}+P'_V)t}  W e^{-i(H_{\bar{V}}+P'_V)t}e^{i(H_{\bar{V}}+P_V)t}\\
=&\frac{d_V}{d\cdot d_W}{\rm Tr}\left[ {\rm Tr}_We^{-i(H_{\bar{V}}+P_V)t}e^{i(H_{\bar{V}}+P'_V)t}\cdot {\rm Tr}_We^{-i(H_{\bar{V}}+P_V)t}e^{i(H_{\bar{V}}+P'_V)t}\right].
\end{aligned}
\end{equation}
We now use again the approximation for the reduced dynamics, i.e.,
\begin{equation*}
    {\rm Tr}_W e^{-i(H_{\bar{V}}+P_V)t} e^{-i(H_{\bar{V}}+P'_V)t} = d_W  e^{-i(H_R+P_V+P_W)t}  e^{i(H_R+P'_V+P_W)t},
\end{equation*}
where $R$ labels the part the total system that excludes the subsystems of $W$ and $V$, Note that a new emerged perturbation $P_W$. Hence,
\begin{equation}
\begin{aligned}
\overline{\text{OTOC}}_{4}
=&\frac{d_Vd_W}{d}  {\rm Tr}\ \left[  e^{-i(H_R+P_V+P_W)t}  e^{i(H_R+P'_V+P_W)t}e^{-i(H_R+P_V+P'_W)t}  e^{i(H_R+P'_V+P'_W)t}\right]\\
=&\langle e^{-i(H_R+P_V+P_W)t}  e^{i(H_R+P'_V+P_W)t}e^{-i(H_R+P_V+P'_W)t}  e^{i(H_R+P'_V+P'_W)t}\rangle_{\beta=0}.
\end{aligned}
\end{equation}
This is a special type of LE, with four loops and the perturbations are local in each loop, i.e., $P_W$ and $P_V$ are local perturbation emerged from the contact with the subsystems of $W$ and $V$, respectively. Let's redefine the unperturbed Hamiltonian as $H = H_R + P'_V+P'_W$ in the above expression, and extract a simplified form of this LE,  i.e.,
\begin{equation}
    M(t)=\langle e^{-i(H+P_2)t}e^{i(H+P_1+P_2)t}e^{-i(H+P_1)t}e^{iHt} \rangle_{\beta=0}.
\end{equation}
In Sec.~\ref{D}, we will present an application of this particular LE for detecting the butterfly velocity.

\subsection{Infinite dimensional generalization}
The previous discussions focus on finite dimensional Hilbert spaces \footnote{For 2pt OTOC, a possible generalization for infinite dimensional Hilbert spaces has been studied in \cite{deMelloKoch:2019rxr}.}. In this section we argue that the OTOC-LE connection can be generalized to infinite dimension. The key ingredient is the Haar integral for unitary operators $U$ on an infinite dimensional Hilbert space, $\int\ d\mu(U) \ U^\dag O U$, where $O$ is a trace-class operator and $\mu$ is the Haar measure. 

Here we consider the right Haar measure, which, by definition, is invariant under transformation $U\rightarrow UV$, i.e., $\mu(UV)=\mu(U)$ for any unitary operator $V$, which implies
\begin{equation}
   \int d\mu(U)\ U^\dag O U = V^\dag \left(\int d\mu(U)\ U^\dag O U \right)V.
\end{equation}
This means that the Haar-averaged operator is proportional to the identity operator ${\rm I}$. 

In finite dimensions, its trace can be computed as
\begin{equation}
\begin{aligned}
         {\rm Tr } \int d\mu(U)\ U^\dag O U = &\int d\mu(U)\ {\rm Tr} \left(O\right).
\end{aligned}
\end{equation}
Haar measure is unique up-to a constant multiplication factor; and the unitary groups on finite dimensional Hilbert spaces have finite measures. This allows us to normalize the Haar measure by choosing $\int d\mu = 1$. Under this convention, the averaged operator has the representation
\begin{equation}
    \int d\mu(U)\ U^\dag O U = \frac{1}{d} {\rm Tr} \left(O\right) {\rm I},
\end{equation}
where $d$ is the dimension of the Hilbert space.

For infinite dimensions, the Haar measure is not normalizable, and hence the averaged operator is not trace-class anymore. However, we are interested in the case where the averaged operator is still bounded (the OTOC takes finite values). In this case, the Haar averaged operator can be fixed as a constant multiplied by the identity, $c(O){\rm I}$. The functional $c$ must be linear and invariant under unitary transformation, i.e., $c(O)=c(U^\dag O U)$. By Riesz representation theorem, it is determined, up-to a multiplication factor, to be the trace, i.e., $c(O)\propto {\rm Tr}(O)$. We have the freedom to remove the pre-factor by absorbing it into the definition of the Haar measure. Under this convention, the desired integral for the Haar average matches precisely with the one in finite dimensions. Once this infinite dimensional Haar integral is evaluated, the OTOC-LE connection follows in the same manner as in the finite dimensional case.

If we average the OTOC over a given group of unitaries $\{U_g\}$, rather than performing the average over all unitary operators with respect to Haar measure, we can firmly say that the OTOC-LE connection holds as well, as long as the group average, up-to a constant multiplication factor which can be removed by re-scaling the measure, gives the same result as the Haar average, namely,
\begin{equation}
    \int dU_g U_g^\dag O U_g \propto \int d\mu_{Haar}(U) U^\dag O U = {\rm Tr }\left(O\right) {\rm I}.
\end{equation}
In other words, the group $\{U_g\}$ is an analog of the unitary 1-design in finite dimensions.

As an example, consider the Heisenberg group $\{U(q_1,q_2)=e^{i(q_1\hat{x}+q_2\hat{p})}\}$, where $\hat{x}$ and $\hat{p}$ are the canonical position and momentum operator, $q_1$ and $q_2$ are real numbers. 

To show that the Heisenberg group is a unitary 1-design, we will need to prove, for any trace-class operator $O$,
\begin{equation}
   \mathcal{D}\equiv \int \frac{dq_1}{2\pi}\int dq_2 U^\dag(q_1,q_2)OU(q_1,q_2) = {\rm Tr} \left(O\right) {\rm I}.
\end{equation}
This is equivalent to showing that the above operator $\mathcal{D}$ in the position representation has elements
\begin{equation}
\begin{aligned}
              \langle x|\mathcal{D}|x'\rangle=&\int \frac{dq_1}{2\pi}\int dq_2 \langle x|U^\dag(q_1,q_2)OU(q_1,q_2)|x'\rangle\\
              =&\int dx_1\int dx_2\int \frac{dq_1}{2\pi}\int dq_2\, \langle x|U^\dag(q_1,q_2)|x_1\rangle\\
              &\quad\quad\quad\quad\langle x_1|O|x_2\rangle\langle x_2|U(q_1,q_2)|x'\rangle\\
              =&\int dx_1\int dx_2\, \delta(x-x')\delta(x_1-x_2)\langle x_1|O|x_2\rangle\\
              =& {\rm Tr} \left(O\right) \delta(x-x').
\end{aligned}
\end{equation}

\subsection{Application I: butterfly velocity} \label{D}
The OTOC is designed as a diagnostic for chaos. For chaotic systems, it decays rapidly and converges to a persistent small value. While for integrable systems the OTOCs typically exhibit oscillatory behaviors, with finite recurrent times.

Another intriguing feature of the OTOC is that it can detect information propagation in systems with higher spatial degree of freedoms. For instance, for a 1-$D$ chaotic spin chain with local interactions, if the operators are chosen as Pauli operators on distinct sites, e.g., $W=\hat{\sigma}^z_i$ and $V=\hat{\sigma}^z_j$, the OTOC does not decay immediately. Rather, it stays constant for a finite amount of time. This is the time for the operator $W(t)$, which is initially local on the $i$'th site, to propagate to the $j$'th site. The propagation is ballistic, with a velocity known as the butterfly velocity \cite{Roberts2016-nd,Nahum2018-ta,Khemani2018-cu,Von_Keyserlingk2018-cf,Parker2019-mr}. This effect is absent for a regular LE, which decays immediately even for local perturbations.

Here, as a first novel application of the OTOC-LE connection, we propose to use a multi-loop LE to probe the butterfly velocity in chaotic systems. As derived in Sec~\ref{sec:localOTOC}. the four-loop LE relating a regular 4 point OTOC with local operators takes the form,
\begin{equation}
    M(t)=\langle\Psi|e^{-i(H+P_2)t}e^{i(H+P_1+P_2)t}e^{-i(H+P_1)t}e^{iHt} |\Psi\rangle,
\end{equation}
Here $P_1$ and $P_2$ are both local perturbations. Regular two-loop LEs of the form $\langle e^{i(H+P_1)t}e^{-iHt} \rangle$ and $\langle e^{i(H+P_1)t}e^{-i(H+P2)t} \rangle$ both show instant decays with no dependence on locality.

We apply the above four-loop LE to study the butterfly velocity of a 1-$D$ spin chain system. The Hamiltonian is given by
\begin{equation}
    H = -J\sum_{i=1}^{N-1}\hat{\sigma}_i^z\hat{\sigma}_{i+1}^z - h_x \sum_{i=1}^N\hat{\sigma}_i^x - h_z \sum_{i=1}^N\hat{\sigma}_i^z,
\end{equation}
where $\hbar$ is set to zero. $1/J$ measures the unit of time. The parameters are fixed as $h_x/J=1.05$ and $h_z/J=0.5$, for which the model is known to be chaotic \cite{Gubin2012-bn,Halpern2017-qu,Alonso:2018faw}. In our simulation of the LE, we chose perturbations as $P_1=g\hat{\sigma}_1^z$, the Pauli operator on the first site, and $P_2=g\hat{\sigma}_n^z$ on the $n$'th site. $g=0.2$ is the strength of the perturbations. The total number of spins is $N=12$.

\begin{figure}
    \centering
    \includegraphics[width=0.45\textwidth]{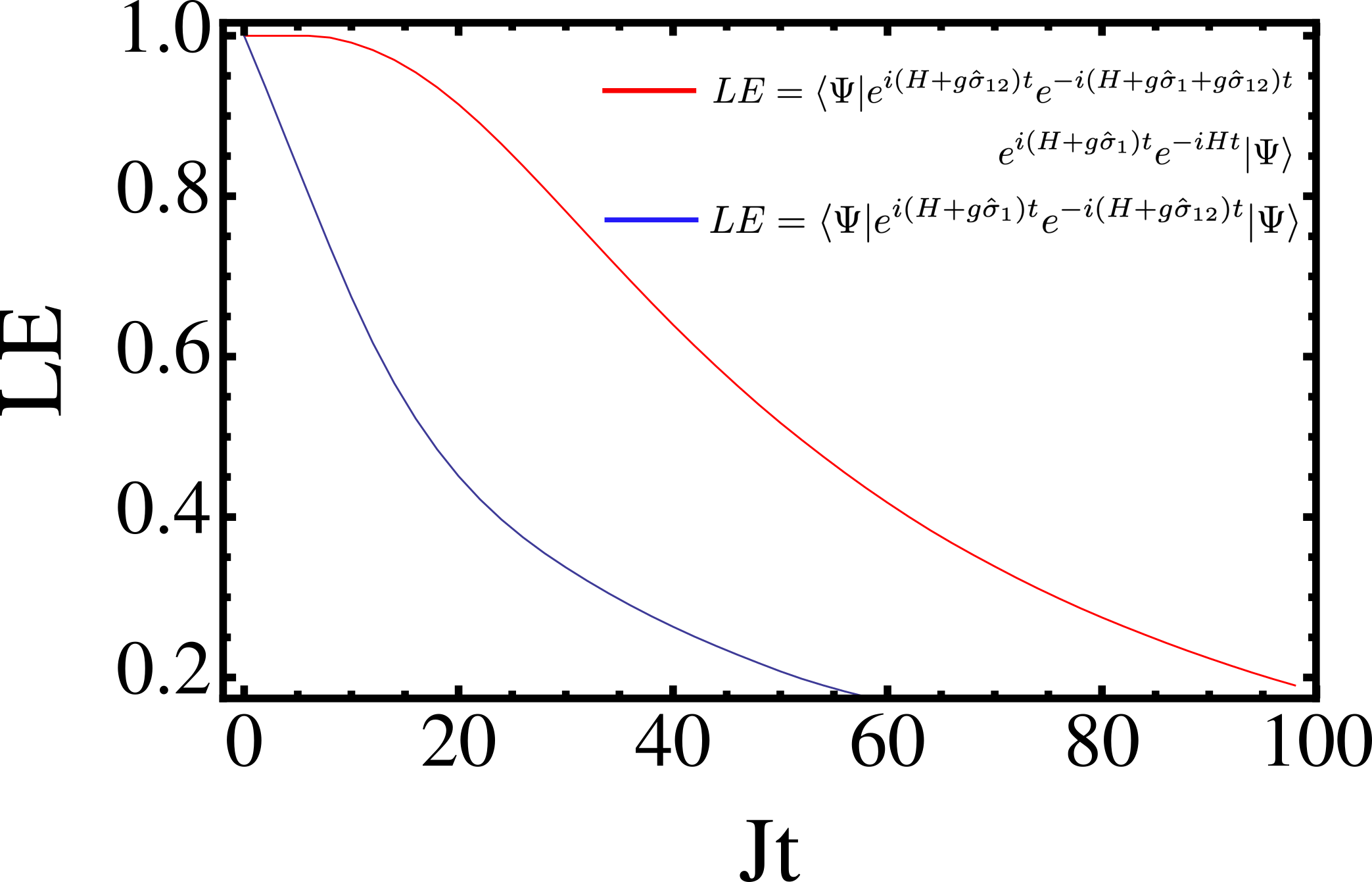}
    \caption{Comparison between the regular two-loop LE and the four-loop LE. The former shows immediate decay after perturbation, while the latter exhibits an initial plateau regime.}
    \label{fig:LE1}
\end{figure}

Fig. \ref{fig:LE1} compares the regular LE, which decays exponentially immediately after perturbation, and the four-loop LE (with $n=12$), which exhibits an initial plateau regime. 

\begin{figure}
    \centering
    \includegraphics[width=0.45\textwidth]{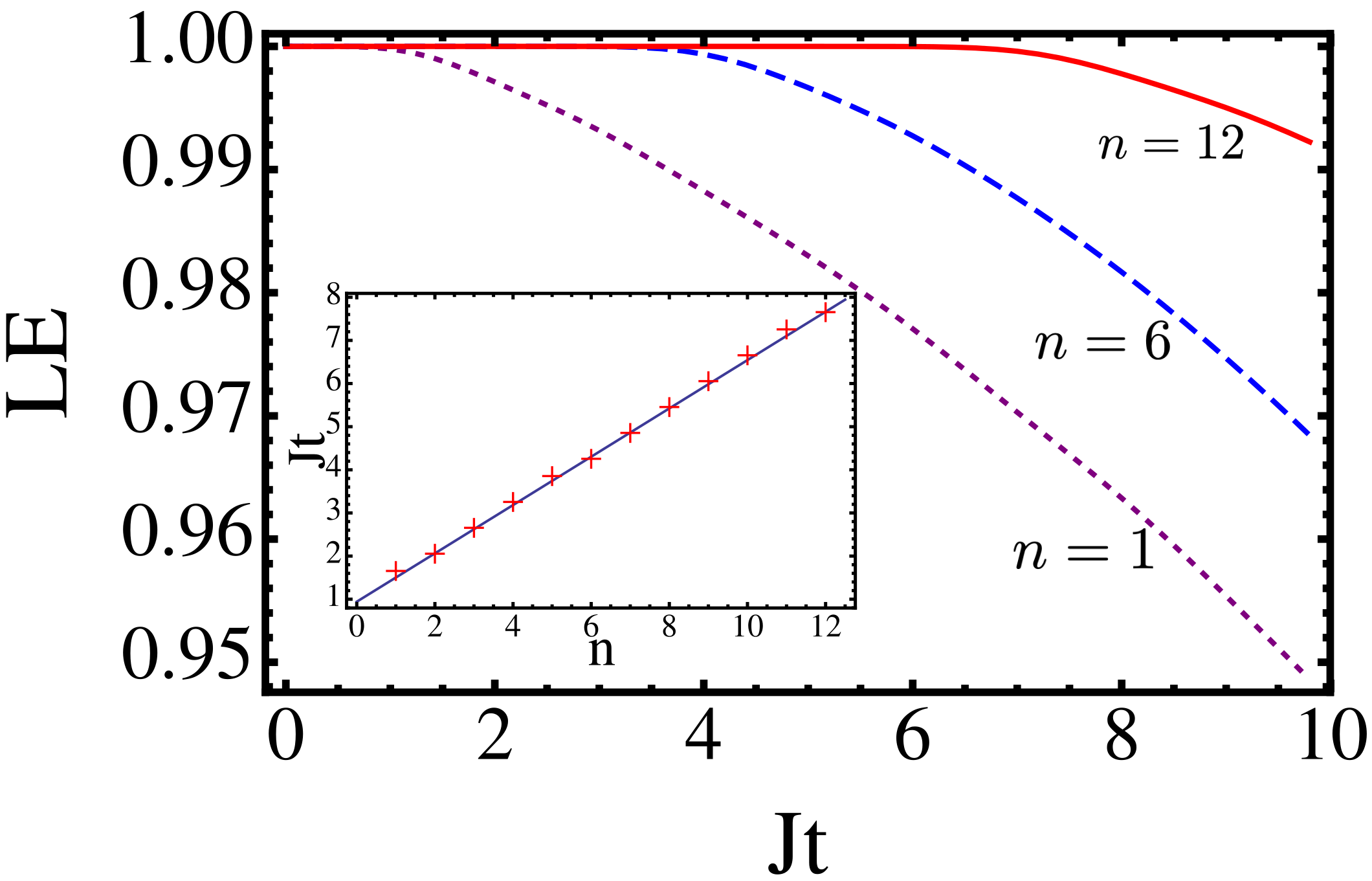}
    \caption{The decay of the four-loop LE at various $n$, where $n$ labels the site on which the second perturbation $P_2$ is applied. The first perturbation $P_1$ is always applied to the first site. Inset: The time at which the LE starts to decay as a function of $n$, the distance between the two sites. Red crosses are numerical data. Solid line is the best liner fit.}
    \label{fig:LE2}
\end{figure}

To extract the butterfly velocity, we simulate the four-loop LE at various $n$'s -- the position of the second perturbation, and read out the time at which the LE starts to decay. In Fig. \ref{fig:LE2}, it can be seen that the larger the distance between the two perturbations, the longer the initial flat regime is; namely, it takes a longer time for the local perturbation to propagate. The width of the plateau regime is proportional to the distance between the two sites (site-$1$ and site-$n$), indicating that the propagation is indeed ballistic.

\subsection{Application II: shockwave and Loschmidt echo}
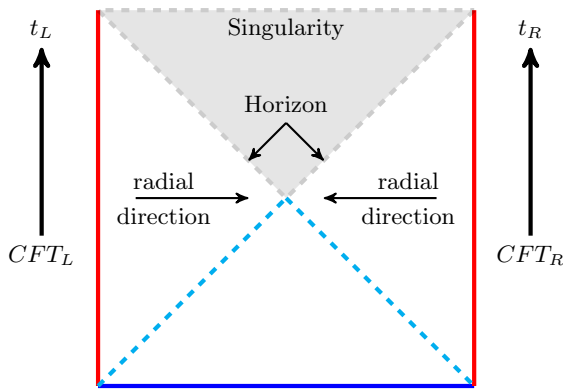
\begin{figure}[ht]
    \centering
\begin{tikzpicture}
\draw[ultra thick, blue] (0.5,0.5) -- (5.5,0.5);
\draw[ultra thick, red] (0.5,0.5) -- (0.5,5.5); 
\draw[ultra thick, red] (5.5,0.5) -- (5.5,5.5);
\draw[ultra thick, cyan, dashed] (0.5,0.5) -- (3,3);
\draw[ultra thick, cyan, dashed] (3,3) -- (5.5,0.5);
 \draw [ultra thick, dashed, draw=black, fill=gray, opacity=0.2]
       (5.5,5.5) -- (3,3) -- (0.5,5.5) -- cycle;
\draw[thick, black, ->] (1,3) -- (2.5,3) node [above,pos=0.25] {radial} node [below,pos=0.25] {direction};
\draw[thick, black, ->] (5,3) -- (3.5,3) node [above,pos=0.25] {radial} node [below,pos=0.25] {direction};
\draw[thick, black, ->] (3,4) -- (3.5,3.5); 
\draw[thick, black, ->] (3,4) -- (2.5,3.5);
\draw[ultra thick, ->] (-0.25,2.5) node[below]{$CFT_L$} -- (-0.25,5) node[above, scale = 1] {$t_L$};
\draw[ultra thick, ->] (6.25,2.5) node[below]{$CFT_R$} -- (6.25,5) node[above, scale = 1] {$t_R$};
\node at (3,4.25) {Horizon};
\node at (3,5.25) {Singularity};  
\end{tikzpicture}
    \caption{Penrose diagram of an eternal AdS-Schwarzschild black hole.}
    \label{fig:AdSS}
\end{figure}

We further illustrate the relation OTOC-LE with a brief application from AdS/CFT correspondence. We shall examine this for an AdS eternal black hole Fig.~\ref{fig:AdSS}. This subsection is primarily based on the following work \cite{stanford2014complexity, chemissany2016holographic,shenker2014multiple,roberts2015localized,susskind2018three,brown2017quantum,brown2018second,swingle2017} and references therein. We shall also shed some light into the possible link that may exist between OTOCs, LE and quantum complexity which we will elaborate on in the next section.

\subsubsection{Echo evolution, Precursors and black holes}
\begin{figure}
    \centering
    \begin{tikzpicture}
\draw[thick] (0,0) -- (0,5.4) -- (5.4,5.4) -- (5.4,0) -- cycle;
\draw [ultra thick, draw=black, fill=gray, opacity=0.2] (0,0) -- (2.7,2.7) -- (0,5.4) -- cycle;
\draw [ultra thick, draw=black, fill=gray, opacity=0.2] (5.4,5.4) -- (2.7,2.7) -- (5.4,0) -- cycle;
\draw[thick,yellow] (0,2.7)--(5.4,2.7);
\draw [thick, yellow] (0,3.6) to [bend right=1] (1.9,3.475);
\draw [thick, red] (1.9,3.475) to [bend right=3] (3.5,3.475);
\draw [thick, yellow] (3.5,3.475) to [bend right=1] (5.4,3.6);
\draw [thick, yellow] (0,4.5) to [bend right=2] (1.1,4.3);
\draw [thick, red] (1.1,4.3) to [bend right=8] (4.3,4.3);
\draw [thick, yellow] (4.3,4.3) to [bend right=2] (5.4,4.5);
\draw [thick, yellow] (0,4.95)--(0.65,4.78);
\draw [thick, red] (0.65,4.78) to [bend right=12] (4.75,4.78);
\draw [thick, yellow] (4.75,4.78)--(5.4,4.95);
\draw [thick, yellow] (0,5.2)--(0.3,5.075);
\draw [thick, red] (0.3,5.075) to [bend right=20] (5.1,5.075);
\draw [thick, yellow] (5.1,5.075)--(5.4,5.2);
\draw [thick, red] (0,5.4) to [bend right=28] (5.4,5.4);
\draw [thick, yellow] (0,1.8) to [bend left=1] (1.9,1.925);
\draw [thick, red] (1.9,1.925) to [bend left=3] (3.5,1.925);
\draw [thick, yellow] (3.5,1.925) to [bend left=1] (5.4,1.8);
\draw [thick, yellow] (0,0.9) to [bend left=2] (1.1,1.1);
\draw [thick, red] (1.1,1.1) to [bend left=8] (4.3,1.1);
\draw [thick, yellow] (4.3,1.1) to [bend left=2] (5.4,0.9);
\draw [thick, yellow] (0,0.45)--(0.65,0.62);
\draw [thick, red] (0.65,0.62) to [bend left=12] (4.75,0.62);
\draw [thick, yellow] (4.75,0.62)--(5.4,0.45);
\draw [thick, yellow] (0,0.2)--(0.3,0.325);
\draw [thick, red] (0.3,0.325) to [bend left=20] (5.1,0.325);
\draw [thick, yellow] (5.1,0.325)--(5.4,0.2);
\draw [thick, red] (0,0) to [bend left=28] (5.4,0);
\node at (-0,2.7)[anchor=east] {$\ket{TFD}$};
\node at (5.4,2.7)[anchor=west] {$t=0$};

    \end{tikzpicture}
    \caption{ The dark red surfaces are the maximal spacelike slices foliating the diagram behind the horizon and representing the wormhole.}
    \label{fig:TwoE}
\end{figure}
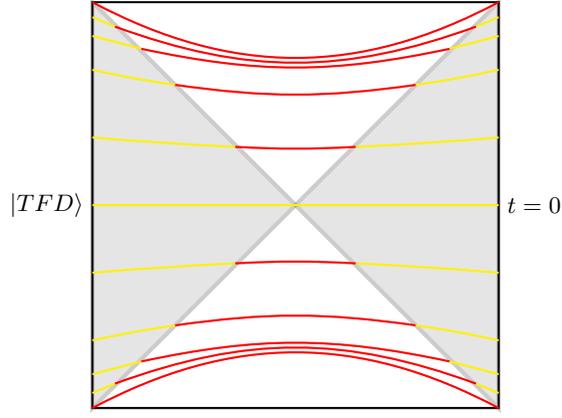

\begin{figure}[ht]
    \centering
\begin{tikzpicture}
\draw[ultra thick, blue] (0,0) -- (5,0);
\draw[ultra thick, blue] (0,5) -- (5,5);
\draw[ultra thick, red] (0,0) -- (0,5); 
\draw[ultra thick, red] (5,0) -- (5,5);
\draw[thick, gray, dashed] (0,0) -- (5,5);
\draw[thick, gray, dashed] (5,0) -- (0,5);
\draw [ultra thick, purple] (0,0.5) -- (2.5,3);
\draw [ultra thick, purple, ->] (2.5,3) -- (4.5,5) node[midway,sloped,above] {\textbf{shockwave}};
\filldraw[black] (0,0.5) circle (2pt) node[anchor=north west] {\large{$t_L = - t_{\omega}$}};
\node[anchor= west] at (5,2.5) {$t=0$};
\node[anchor= east] at (0,2.5) {$t=0$};
\draw[cyan,->,thick] (-0.5,0.7)--(-0.4,1.45);
\draw[cyan,thick] (-0.4,1.45)--(-0.3,2.2);
\draw[cyan,thick] (-0.5,0.7)--(-0.6,1.45);
\draw[cyan,<-,thick] (-0.6,1.45)--(-0.7,2.2);
\end{tikzpicture}
    \caption{The operator $W_{L}$ creates an infalling quantum at $|t_{w}|>> t_{*},$ where $t_{*}$ is the scrambling time. It undergoes a huge blue shift as it moves at the speed of light toward the horizon.}
    \label{fig:tltw}
\end{figure}
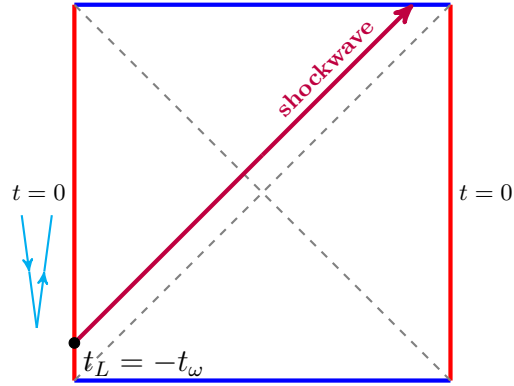

Let's consider two entangled black holes connected by an Einstein-Rosen bridge, aka wormhole Fig.\ref{fig:TwoE}. The holographic description of the wormwhole volume is quantified by the complexity of the quantum state of the dual pair of CFTs at time $t.$  For a given thermofield double state, we can evolve, for instance,  the left side back in time for a time $\Delta t_L = -t_w$ and then apply a simple localized precursor perturbation $W_L$ that adds a thermal quantum; a localized packet of energy in the left side. Then, we evolve this state forward in time, $\Delta t_L = t_w$ (see Fig.\ref{fig:tltw}). Due to the fact that the quantum state loses its memory,  the left target state has to differ from the left initial state. 

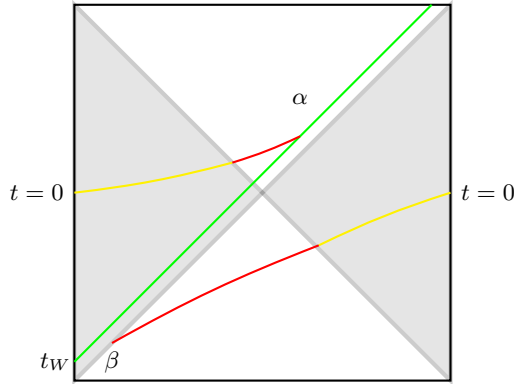
\begin{figure}
    \centering
    \begin{tikzpicture}
\draw[thick] (0,0) -- (0,5) -- (5,5) -- (5,0) -- cycle;
\draw [ultra thick, draw=black, fill=gray, opacity=0.2] (0,0) -- (2.5,2.5) -- (0,5) -- cycle;
\draw [ultra thick, draw=black, fill=gray, opacity=0.2] (5,5) -- (2.5,2.5) -- (5,0) -- cycle;
\draw [thick ,green] (0,0.25) -- (4.75,5); 
\draw [thick,yellow] (0,2.5) to [bend right=4] (2.1,2.9);
\draw [thick, red] (2.1,2.9) to [bend right=4] (3,3.25);
\draw [thick,red] (0.5,0.5) to [bend left=4] (3.25,1.8);
\draw [thick, yellow] (3.25,1.8) to [bend left=4] (5,2.5);
\node at (-0.25,0.25) {$t_W$};
\node at (-0.5,2.5) {$t=0$};
\node at (5.5,2.5) {$t=0$};
\node at (3,3.75) {$\alpha$};
\node at (0.5,0.25) {$\beta$};
    \end{tikzpicture}
    \caption{In the absence of the shockwave the size of the maximal spacelike slice formed behind the horizon is null (it goes through the intersection point of the bifurcate horizon). Adding the shockwave allows the maximal slice to acquire a considerable volume as indicated by the red surfaces.   }
     \label{fig:SingleS}
\end{figure}

In AdS spacetime, this energy source starts to warp the spacetime near the horizon by creating a gravitational shockwave which expands away from the source, and remains highly energetic for most of its worldline. This kick changes the geometry and leads to a larger wormhole compared to the initial one. More precisely,  the wormhole owes its growth to the gravitaional back-reaction on the shape of the geometry (or alternatively to the hop and displacement of the trajectories crossing the shockwave Fig.\ref{fig:SingleS}). This corresponds, on  the CFT side, to a greater decay for the correlation between the two different sides of the thermofield-double state.  Consequently, one infers that there exists a correspondence between the amount of energy the shockwaves produce and the rate of decay of the correlations. In what follows, we shall provide a rough picture as to how this might play out. 
\subsubsection{Single shock}
Let us apply the \emph{echo} evolution $e^{-iH_L t_w}W_L e^{iH_L t_w}$ to the thermofield-double state:
\begin{eqnarray} \ket{w}&=&e^{-iH_L t_w}W_L e^{iH_L t_w}\ket{TFD}
=W_{L}(t_{\omega}) \ket{TFD}\end{eqnarray} 
where for any single sided operator (precursor) $W$, $W_L=W\otimes\mathbb{I}$ and $W_R=\mathbb{I}\otimes W$. The operator $W_{L}(t_{\omega})$ is a Schr\''{o}dinger picture operator acting at time $t=0.$ The effect of $W_{L}(t_{\omega})$ amounts to adding at $t_\omega$ a thermal quantum to the left side. Note that despite the fact that the thermal quantum, being localized low energy perturbation, created by $W_{L}(t_{\omega}),$ feeds a tiny bit of energy to the black hole, for most of its worldline it is astronomically energetic shockwave.

The two-sided correlator is found to be (see for instance \cite{swingle2017})
 \begin{equation}
 \begin{aligned}
     \bra{w}V_L\otimes V_R^T\ket{w} 
   & =\bra{TFD}e^{-iH_L t_w}W_L^{\dagger} e^{iH_L t_w}V_L \otimes V_R^T e^{-iH_L t_w}W_L e^{iH_L t_w}\ket{TFD},\\
    =&\bra{TFD}W_L^{\dagger}(-t_w) V_L \otimes V_R^T W_L(-t_w) \ket{TFD},\\
    =&\bra{TFD}\left(W_L^{\dagger}(-t_w)\otimes\mathbb{I}\right)\left(V_L \otimes V_R^T\right)(W_L(-t_w)\otimes\mathbb{I}) \ket{TFD},\\
    =&\bra{TFD}W_L^{\dagger}(-t_w)V_L W_L(-t_w)\otimes V_R^T\ket{TFD}.
\end{aligned}
\end{equation}
The transpose ``T" is in the energy basis. The state is subject to the so-called operator pushing property by which we mean 
 \begin{equation}\hat{W}_R\ket{max}=\hat{W}_L^T\ket{max}\label{pushing}\end{equation}
 with $\ket{max}$ being a maximally entangled state. Using (\ref{pushing}), one can push $V^{T}$ from the right to the left. One therefore can convert the two-sided correlator to a one-sided correlator, i.e., 
 \begin{eqnarray}\label{twotoone}
    \underbrace{\bra{w}V_L\otimes V_R^T\ket{w}_{\beta=0}}_{\substack{\text{correlation between the}\\ \text{2 sides after perturbation}}} &=& \underbrace{\braket{W_L^{\dagger}(-t_w)V_L W_L(-t_w)V_L}_{\beta=0}}_{\substack{\text{4-point OTOC at}\\ \text{time}=-t_w}}\nonumber\\
    &=& \frac{1}{2^{N}} \textrm{Tr}(W_L^{\dagger}(-t_w)V_L W_L(-t_w)V_L).\nonumber\\
\end{eqnarray}
The negative time is not profoundly significant. Evolving the system according to negative time is expected to have the same behaviour as evolving it with positive time. As a matter of fact, the behaviour of the OTOC can be generic for different local operators. In this case for any thermal state the aforementioned argument may be extracted from $\braket{W^{\dagger}(-t_w)VW(-t_w)V}_{\beta}=\braket{W^{\dagger}V(t)WV(t)}_{\beta}$. The generalization of the previous claim to general temperature, for which the obtained OTOCs are thermally regulated, is straightforward \cite{swingle2017}. 

\subsubsection{Multiple shocks}
The lesson one can draw out from the previous single shock case is that the more shockwaves (energy) you feed the black hole with, the greater the decay of the OTOCs becomes. To create two shockwaves one needs to consecutively repeat the process introduced above twice, that is,
\begin{equation}\begin{split}
 \ket{w_{1,2}} &=e^{-iH_Lt_2}W_2e^{iH_Lt_2}e^{-iH_Lt_1}W_1e^{iH_Lt_1}\ket{TFD},\\
&= W_2(-t_2)W_1(-t_1)\ket{TFD},\\
&=W_{multi}(t_{2},t_{1})\ket{TFD}.
\end{split}\end{equation}
from which one can derive
\begin{equation}
    \underbrace{\bra{w_{1,2}}V_L\otimes V_R^T\ket{w_{1,2}}}_{\substack{\text{correlation between 2}\\ \text{sides after perturbing}\\ \text{twice}}}\\
= \underbrace{\braket{W_1^{\dagger}(-t_1)W_2^{\dagger}(-t_2)V_LW_2(-t_2)W_1(-t_1)V_L}}_{\text{6-point OTOC}}.
\end{equation}
Notice that we can get rid of the negative time because of the same previously mentioned reasons. This makes our claim true for the case of two shockwaves.
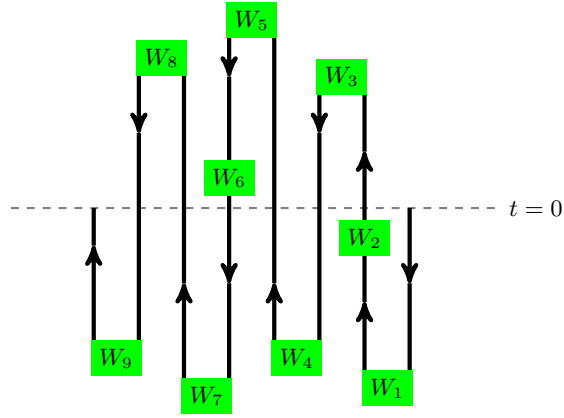
\begin{figure}[ht]
    \centering
\begin{tikzpicture}
\draw[thick, dashed, gray] (0,0)--(6.5,0) node[anchor=west] {\textcolor{black}{$t=0$}}; 
\draw[ultra thick] (1.1,0)--(1.1,-0.5);
\draw[ultra thick, <-] (1.1,-0.5)--(1.1,-2);
\draw[ultra thick, <-] (1.7,1)--(1.7,2);
\draw[ultra thick] (1.7,-2)--(1.7,1);
\draw[ultra thick, <-] (2.3,-1)--(2.3,-2.5);
\draw[ultra thick] (2.3,-1)--(2.3,2);
\draw[ultra thick, ->] (2.9,2.5)--(2.9,1.75);
\draw[ultra thick, ->] (2.9,1.75)--(2.9,-1);
\draw[ultra thick] (2.9,-1)--(2.9,-2.5);
\draw[ultra thick, ->] (3.5,-2)--(3.5,-1);
\draw[ultra thick] (3.5,-1)--(3.5,2.5);
\draw[ultra thick, ->] (4.1,1.75)--(4.1,1);
\draw[ultra thick]  (4.1,1)--(4.1,-2);
\draw [ultra thick, ->] (4.7,-2.4)--(4.7,-1.25);
\draw[ultra thick, ->] (4.7,-1.25)--(4.7,0.75);
\draw[ultra thick] (4.7,0.75)--(4.7,1.75);
\draw[ultra thick, ->] (5.3,0)--(5.3,-1);
\draw[ultra thick] (5.3,-2.4)--(5.3,-1);
\node (rect) at (1.4,-2) [fill, green,thick,minimum width=0.4cm,minimum height=0.4cm] {\textcolor{black}{$W_9$}};
\node (rect) at (2,2) [fill, green,thick,minimum width=0.4cm,minimum height=0.4cm] {\textcolor{black}{$W_8$}};
\node (rect) at (2.6,-2.5) [fill, green,thick,minimum width=0.4cm,minimum height=0.4cm] {\textcolor{black}{$W_7$}};
\node (rect) at (2.9,0.4) [fill, green,thick,minimum width=0.4cm,minimum height=0.4cm] {\textcolor{black}{$W_6$}};
\node (rect) at (3.2,2.5) [fill, green,thick,minimum width=0.4cm,minimum height=0.4cm] {\textcolor{black}{$W_5$}};
\node (rect) at (3.8,-2) [fill, green,thick,minimum width=0.4cm,minimum height=0.4cm] {\textcolor{black}{$W_{4}$}};
\node (rect) at (4.4,1.75) [fill, green,thick,minimum width=0.4cm,minimum height=0.4cm] {\textcolor{black}{$W_{3}$}};
\node (rect) at (4.7,-0.4) [fill, green,thick,minimum width=0.4cm,minimum height=0.4cm] {\textcolor{black}{$W_{2}$}};
\node (rect) at (5,-2.4) [fill, green,thick,minimum width=0.4cm,minimum height=0.4cm] {\textcolor{black}{$W_{1}$}};
\end{tikzpicture}
  \caption{Multifold echo with $2(k-1)$ loops. Each green insertion represents a tiny perturbation. The arrows point toward the order in which the precursors $W_{i}$'s apply.}
     \label{fig:MultiS}
\end{figure}
 Upon the application of multiple $(k-1)$ operators $W(t)$  on $\ket{TFD}$ (pictorially presented in Fig.\ref{fig:MultiS}), One can write
\begin{equation}\ket{w_{1,2,...,k-1}}=W_{multi}(t_{k-1},\cdots,t_{1})\ket{TFD},\end{equation}
from which one ought to obtain $2k$-OTOC where $k>1,$ i.e., 
\begin{equation}
        \underbrace{\bra{w_{1,2,...,k-1}}V_L\otimes V_R^T\ket{w_{1,2,...,k-1}}}_{\substack{\text{correlation between 2 sides}\\ \text{ after perturbing (k-1) times}}} \\
        =
   \underbrace{\braket{W_1^{\dagger}(-t_1)...W_{k}^{\dagger}(-t_k)V_LW_k(-t_k)...W_1(-t_1) V_L}}_{\text{2k-point OTOC}}.
\end{equation}

\subsubsection{The triangle links}
As claimed above the shockwave has a large effect on the geometry. Without the shockwaves the volume of the maximal slice behind the horizon at $t=0$ is null (it goes through the bifurcate horizon). Upon the creation of the shockwaves the spatial maximal slice, representing the wormhole connecting the two-sided entangled black holes, gains a significant volume. Roughly speaking, one can anticipate that the correlation exponentially decays with the size (length) $L(t)$ of the wormhole. Thus \footnote{We assume that the perturbations separately and successively act such that effect of $W_{i}$ fills out the entire system before $W_{i+1}$ kicks in.},
\begin{equation} 
\bra{w}V_L\otimes V_R^T\ket{w}_{\beta=0}\sim e^{-\frac{L(t)}{l_{\text{\text{Ads}}}}}.
\end{equation}
Using (\ref{twotoone}), it yields 
\begin{equation} \braket{W_L^{\dagger}(-t_w)V_L W_L(-t_w)V_L}_{\beta=0}\sim e^{-\frac{L(t)}{l_{\text{Ads}}}}\label{ladsC}.\end{equation}
From the OTOC-LE connection, we have
\begin{equation}
    \int\limits_{Haar}dWdV\langle W_L^{\dagger}(t)V_L^{\dagger}W_L(t)V_L\rangle_{\beta=0}
    \approx |\langle e^{iH_L t}e^{-i(H_L+\Delta)t}\rangle|^2.
\end{equation}
This implies, 

\begin{equation}
 \bigg|\bra{TFD}e^{iH_Lt_w}W_Le^{-iH_Lt_w}\ket{TFD}\bigg|^2 = \bigg|\braket{TFD|w}\bigg|^2.
\end{equation}
We find that \begin{equation}
    \bigg|\braket{TFD|w}\bigg|^2=\int\limits_{Haar}dWdV e^{-l(t)/l_{\text{AdS}}}.\quad  
\end{equation} 
It was conjectured by Susskind et.al. that quantum complexity $\mathcal{C}$ (precisely introduced in section III) is related to the size (length/volume) $\mathcal{V}$ of the wormhole connecting the two entangled black holes, i.e., $\mathcal{C}=\frac{\mathcal{V}}{G l_{AdS}}.$ Combining all these together we end up with 
\begin{equation}
e^{-iH_{L}t_{w}}W e^{iH_{L}t_{w}}\ket{TFD}= \ket{w}. \label{new3} \\
\end{equation}
Then conjugating (\ref{new3}) both sides by $\bra{TFD}$ we get,
\begin{align}
    \begin{split}
\bra{TFD}e^{iH_{L}t_{w}}e^{-i(H_{L}+V)t_{w}}\ket{TFD}=\langle TFD|w\rangle.
\end{split}
\end{align}
Finally we arrive at the following,
\begin{equation}
\begin{aligned}
&|\bra{TFD}e^{iH_{L}t_{w}}e^{-i(H_{L}+V)t_{w}}\ket{TFD}|^2\\
=& |\langle TFD|w\rangle|^2\sim LE\sim  e^{-L(t)/l_{\text{AdS}}}=e^{-\mathcal{C}}.
\end{aligned}
\end{equation}

This derivation involves only one single shockwave. However, one can incorporate multitude of shocks for which the one-fold LE is superseded by multi-fold LE and the complexity associated with one localized precursor is replaced by $\mathcal{C}[W_{multi}(t_{k-1},\cdots, t_{1})]$ such that
\begin{eqnarray} 2k\textrm{-OTOC}&\sim&e^{\frac{\tilde{L}(t)}{\l_{AdS}}}\\
 2k\textrm{-OTOC}&=&LE_{multi}\sim e^{-\mathcal{C}_{multi}},\end{eqnarray}
 where $\tilde{L}(t)$ is the stretched length of the ERB (wormhole) behind the horizon.

\subsubsection{Perspectives from infinite dimensional continuous variable systems}
A relation between an operator's distribution in phase space and OTOCs in continuous variable (CV) system has been established in \cite{zhuang2019scrambling}. Consider an operator that spreads in phase space having width/volume $\mathcal{V}.$ The OTOC was found to be \cite{zhuang2019scrambling}
\begin{equation} \mathcal{C}_{2}(\pmb{\xi}_{1},\pmb{\xi}_{2};t)_{\rho}\sim e^{-\mathcal{V}|\pmb{\xi}_{2}|^2}.\label{OTV}\end{equation}
To derive (\ref{OTV}) we shall introduce a few definitions and quantities. We begin by defining the displacement operator, the analog of the Pauli operator in discrete variables, for a simple harmonic oscillator (single mode CV system)
\begin{equation} D(\xi_{2},\xi_{2})\equiv e^{[i(\xi_{2}q-\xi_{1}p)]}.\end{equation}
Such shifts operators, being e.g., elements of the Heisenberg group, form a complete basis and act on a coherent state in phase space. For $N$-mode CV system they read
\begin{equation} D(\pmb{\xi})=e^{i \pmb{x}^{T}\pmb{\Omega} \pmb{\xi}},\quad \pmb{\Omega}=\oplus^{N}_{k=1}\begin{pmatrix}0&1\\-1&0\end{pmatrix},\quad \pmb{\xi}\in R^{2N}, \end{equation}
with $\pmb{x}=(q_1,q_{2},\cdots,q_{N},p_{N})$ being the vector of quadrature operators. These $N$-mode displacement operators satisfy
\begin{eqnarray} &&\textrm{Tr}(D(\pmb{\xi})D(\pmb{\xi}'))=\pi^{N}\delta(\pmb{\xi}+\pmb{\xi}'),\\
&&\frac{1}{\pi^{N}}\int d^{2N}\pmb{\xi} D(\pmb{\xi})A D^{\dagger}(\pmb{\xi})=\textrm{Tr}(A)\,\pmb{I}.\end{eqnarray}

The CV OTOC is defined to be
\begin{equation} \mathcal{C}_{2}(\pmb{\xi}_{1},\pmb{\xi}_{2};t)_{\rho}=\textrm{Tr}\left[\rho D^{\dagger}(\pmb{\xi}_{1};t)D^{\dagger}(\pmb{\xi}_{2})D(\pmb{\xi}_{1};t)D(\pmb{\xi}_{2})\right]\end{equation}
where the so-called displacement operator takes the following form 
\begin{eqnarray} D(\pmb{\xi}_{2};t)&=&\frac{1}{\pi^{N}}\int d^{2N}\pmb{\xi}_{2}\, \chi[\pmb{\xi}_{2};D(\pmb{\xi}_{1};t)]D(-\pmb{\xi}_{2})\\
D(\pmb{\xi}_{1};t)&\equiv& U(t)^{\dagger} D(\pmb{\xi}_{1})U(t).\end{eqnarray}
From the above decomposition, which is allowed by the the completeness of displacement operators, one can infer that scrambling in the CV system is featured by the growth of the Wigner characteristic $\chi[\pmb{\xi}_{2}; D(\pmb{\xi}_{1};t)]$ given by 
\begin{eqnarray}\chi(\pmb{\xi};A)&\equiv& \textrm{Tr}[A D(\pmb{\xi})],\\
\chi[\pmb{\xi}_{2};D(\pmb{\xi}_{1};0)]&=&\pi^{N}\delta(\pmb{\xi}_{2}+\pmb{\xi}_{1}).\end{eqnarray}
Now, given 
\begin{equation} \chi[\pmb{\xi}; D(\pmb{\xi}_{1};t)]\sim e^{\frac{|\pmb{\xi}-\pmb{\xi}_{1}|^{2}}{2\mathcal{V}}}\end{equation} and making use of the formulae presented above leads to (\ref{OTV}).  This shows that a larger width of the phase space results in greater decay of the OTOC. One should be able to directly relate the increase of the operator volume in the phase space with the size of the wormhole in the two entangled black holes model studied above rendering the connection between OTOC-LE and complexity more rigorous. This correspondence may be achieved by matching the norm of the displacement vector with the AdS radius, i.e.,  $|\xi|\sim 1/l_{\text{AdS}}$ (cf. eq.(\ref{ladsC})).

It is worth emphasising that averaging the OTOCs over ensembles of displacement operators may enable us to measure a coarse-grained spread of a time-evolved operator in phase space. This may allow one to gain better understanding into the link between various diagnostics. For more about using the average OTOCs as probes for finer-grained aspects of operator distribution, we refer the reader to \cite{zhuang2019scrambling}. In section III we shall pursue a slightly different path to establish such a connection between the three diagnostics. 

Due to some very recent progress one is now able to test these predictions. Based on the formalism developed in \cite{zhuang2019scrambling} and a generalization of quantum teleportation mechanism, a detailed experimental blueprint has been put forward \cite{schuster2021many}. The proposed experimental protocols can be potentially generalized to include (multi)-shockwaves. Along similar lines, Brown et al. have recently set a long-term goal of studying models of quantum gravity in the lab \cite{brown2019quantum,nezami2021quantum}, which could mimic in particular the two entangled black hole set-up. More precisely, they put forward holographic teleportation protocols that can be readily executed in table-top experiments. 
These quantum-teleportation-inspired experimental protocols are malleable to be devised such that they could potentially include multiple shockwaves, from which one would be able to test our predictions in experiments.

In summary, chaos is a keynote ingredient intimately related to the onset of thermalization. It has been shown that a signal of chaos is encoded in the behaviour of OTOCs and exponential growth of the commutators. In the gravity side, this growth manifests itself as a near horizon higher energy scattering/collision; semi-classically controlled by a shockwave geometry.  
A particle propagating at the speed of light can be described as a null-like delta function of the stress-energy tied to the horizon of the black hole. This source triggers what is called a gravitational shockwave that moves far away from the source. Going through the shockwave results in a kick in a certain null-like direction. Such an extra kick modifies the geometry, leading to a kind of decorrelation between the two sides of the perturbed thermofield double when the kick becomes strong. The boost the particle has gone through after falling for a time $t_{\omega}$ is proportional to the strength of the kick that scales like the proper energy $\sim e^{2\pi t_{\omega}/\beta}$. This dependence on the exponential time is one quantifier/measure of quantum chaos in the given system, with $\lambda_{L}=\frac{2\pi}{\beta}$--being the quantum Lyapunov exponent of the black hole. This picture can be generalized for multiple shockwaves, which implies a clear connection between quantum chaos/fast scrambling and the recently established transversability of the wormhole.


\section{Loschmidt echo and complexity}
We will start with a brief review of the circuit complexity by using the Nielsen's method \cite{NL1,Jefferson:2017sdb}. 
Given a reference state $|\psi_{s=0}\rangle$, a target state $|\psi_{s=1}\rangle$, and a set of elementary gates $\{\text{exp}(-i\,M_{I})\}$ where the $\{M_I\}$ are
group generators, the goal is to build the most efficient circuit $U (s)$ that starts at the reference state and terminates at the target state:
\begin{equation}
    |\Psi_{s=1}\rangle = U (s=1) |\Psi_{s=0}\rangle,
\end{equation}
where $U (s)$ is the path-ordered operator
\begin{equation}
U(s)= {\overleftarrow{\mathcal{P}}} \exp[- i \int_0^{s} \hspace{-0.1in} ds' H(s') ] \ ,
\end{equation}
where $H(s')$ is the Hamiltonian and can be written as
\begin{equation}
    H(s)= Y(s)^{I} M_{I}\, .
\end{equation}
The coefficients $Y^I$ are the control functions that dictates which gate will act at a given value of the parameter. The control function is basically a tangent vector in the space of unitaries and satisfy the Schrodinger equation
\begin{equation}
\frac{d U(s)}{ds} = -i\, Y(s)^I M_I U(s)\,.
\end{equation}
Then we define a cost functional $\mathcal{F (U, \dot U)}$ as follows:
\begin{equation}
{\mathcal C}(U)= \int_0^1 \mathcal{F} (U, \dot U) ds\, .
\end{equation}
Minimizing this cost functional gives us the optimal circuit. There are different choices for the cost functional \cite{Jefferson:2017sdb}. In this paper we will consider
\begin{equation}
\mathcal{F}_2 (U, Y)  = \sqrt{\sum_I (Y^I)^2}\, .
\label{quadCost}
\end{equation}

\subsection{Introducing complexity}
Recently complexity has been demonstrated as an equally powerful and computationally simpler quantity in some cases than OTOC to diagnose the chaotic behaviour of a quantum system \cite{Ali2019-ks, Cotler:2017jue}. Since all three of these quantities--LE, OTOC and Complexity--are providing similar information about the chaotic system, it is natural to anticipate that these three quantities are related to each other. In the previous sections we have established that the sub-system LE and averaged OTOC are the same. Therefore, to establish the relationship between the three quantities, we only need to explore the connection between LE and complexity. 

To make progress in this direction, we will use the complexity for a particular quantum circuit from the inverted oscillator model:
\begin{equation*}
    H = \frac{1}{2}p^2 + \frac{\Omega^2}{2}x^2,\ \text{where}\ \Omega^2=m^2-\lambda.
\end{equation*}
Classically, the inverted harmonic oscillator has an unstable fixed point and is not a chaotic system in the strict sense. Nonetheless, it has been used as a powerful toy model for studying quantum chaos in various quantum field theories \cite{Blume, Morita2019-de,Bueno2019-zh,Betzios:2016yaq,Hegde:2018xub}, mostly because it is an exactly solvable system.
The oscillator can be tuned to the regular and chaotic regime by changing the value of $\lambda$, i.e., for $\lambda<m^2$ the oscillator is simple, while for $\lambda>m^2$ the oscillator is inverted and chaotic.

In \cite{Ali2019-ks}, it was shown that the appropriate quantum circuit in this regard is the one where the target state $|\psi_2\rangle$ is obtained by evolving a reference state $|\psi_0\rangle$ forward in time by Hamiltonian $H$ and then backward in time with slightly different Hamiltonian $H+\delta H$ as follows 
\begin{equation} \label{target}
   | \psi_2\rangle = e^{i (H+\delta)  t}e^{-iH t} |\psi_0\rangle .
\end{equation}
 For the inverted harmonic oscillator model the authors in \cite{Ali2019-ks} showed that the complexity evaluated by using the covariance matrix method \cite{Hackl:2018ptj,Camargo:2018eof} for the above mentioned target state with respect to the reference state $|\psi_0\rangle$ is given by 
\begin{equation}
\mathcal{C} =\frac{1}{2}\left[\cosh^{-1}\left(\frac{\omega_r^2+|\hat \omega(t)|^2}{2\,\omega_r\,\text {Re}  (\hat \omega(t))}\right)\right],
\end{equation}
where $\hat \omega(t)$ is the frequency of the doubly evolved Gaussian target state which has the following form
\begin{equation}
\psi_2(x,t) =\mathcal{\hat N}(t) \exp \left[ -\frac{1}{2}\hat \omega(t) x ^2  \right] \,,
\end{equation}
and
 \begin{equation}
 \label{omedef} \hat \omega(t)= \left[ i \ \Omega' \cot (\Omega' t)  + \frac{\Omega'^2}{  \sin^2 (\Omega' t) (\omega(t) + i \, \Omega' \cot (\Omega' t))}\right]\,. 
 \end{equation}
In the last expression, $\Omega'= \sqrt{m^2 - \lambda'}$ is the frequency associated with the perturbed/slightly different Hamiltonian $H'=\frac{1}{2} p^2 + \frac{\Omega'^2}{2} x^2$ and $\lambda'=\lambda+\delta\lambda$ with $\delta\lambda$ very small. We make this perturbation by hand. 

\begin{figure}[t] 
\centering
\includegraphics[width=0.45\textwidth]{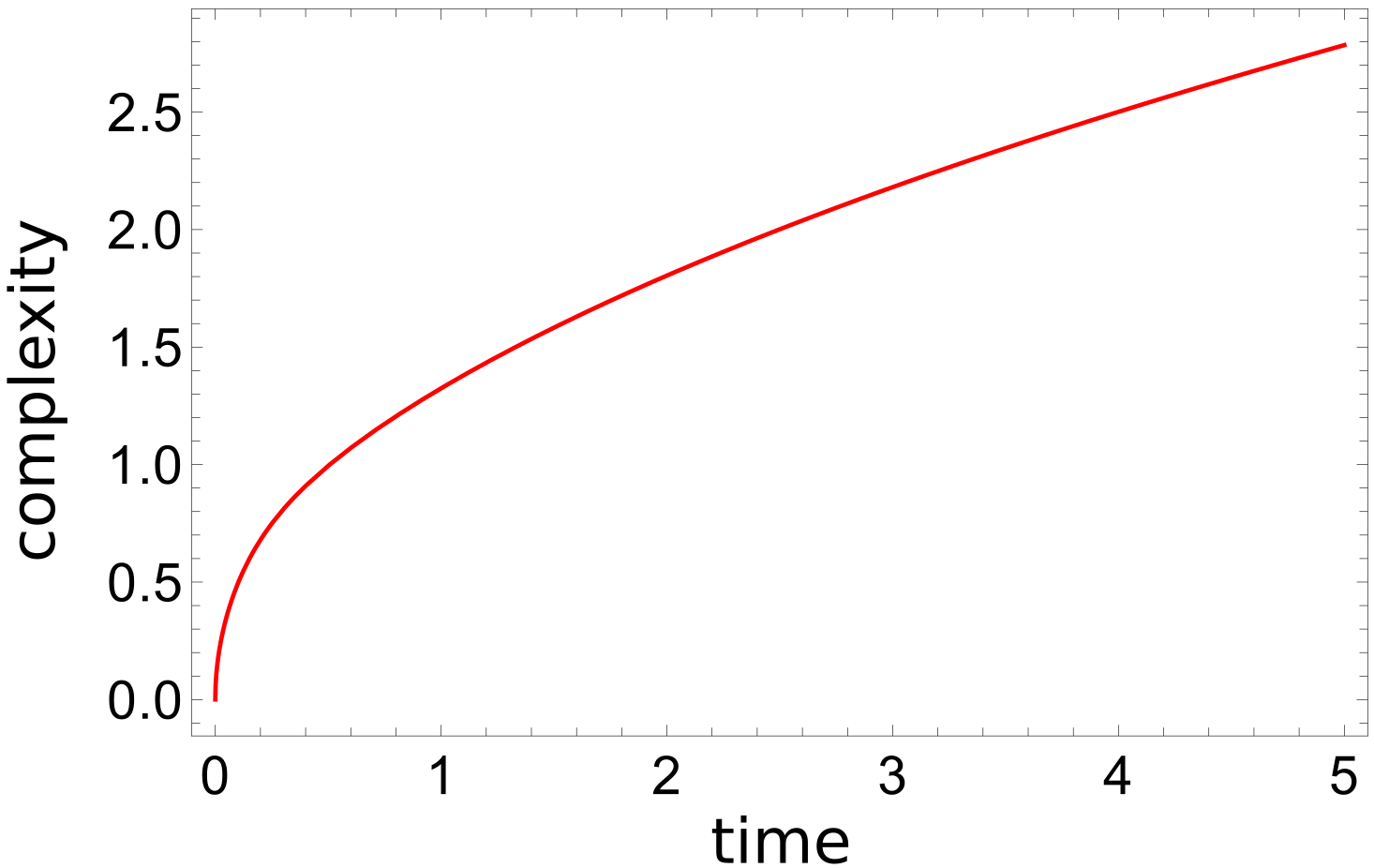}
\caption{Complexity of single time evolved target state for inverted oscillator ($m=1,\lambda=20$).}
\label{single}
\end{figure}

Note that the quantum circuit involving two time-evolutions with slightly different Hamiltonians is crucial for extracting the chaotic nature of the quantum system. Complexity for any target state will not capture similar information as OTOC. For example, the complexity of a target state which is forward evolved only once will not capture the scrambling time for the chaotic system as illustrated in Fig.~\ref{single}. However, there is an alternative quantum circuit that will have the same complexity when evaluated by the covariance matrix method. In that circuit both the reference and target states are basically evolved states but with slightly different Hamiltonians from some other state. Once again this particular circuit also involves two evolutions.

\subsection{LE-complexity connection}
It was shown in \cite{Ali2019-ks} that complexity of the above mentioned target state (\ref{target}) can capture equivalent information such as scrambling time and Lyapunov exponent as the OTOC for an inverted oscillator. In this paper, we want to make this statement more precise by using the fact that averaged OTOC is the same as (very close to) the sub-system LE. In section \ref{D} of this paper, we have proved this for the Heisenberg group. In the current section, we will use an explicit example from the Heisenberg group, namely the inverted oscillator to demonstrate that LE for the full system and complexity are very close quantities.

\begin{figure*}[t] 
\centering
\includegraphics[width=0.9\textwidth]{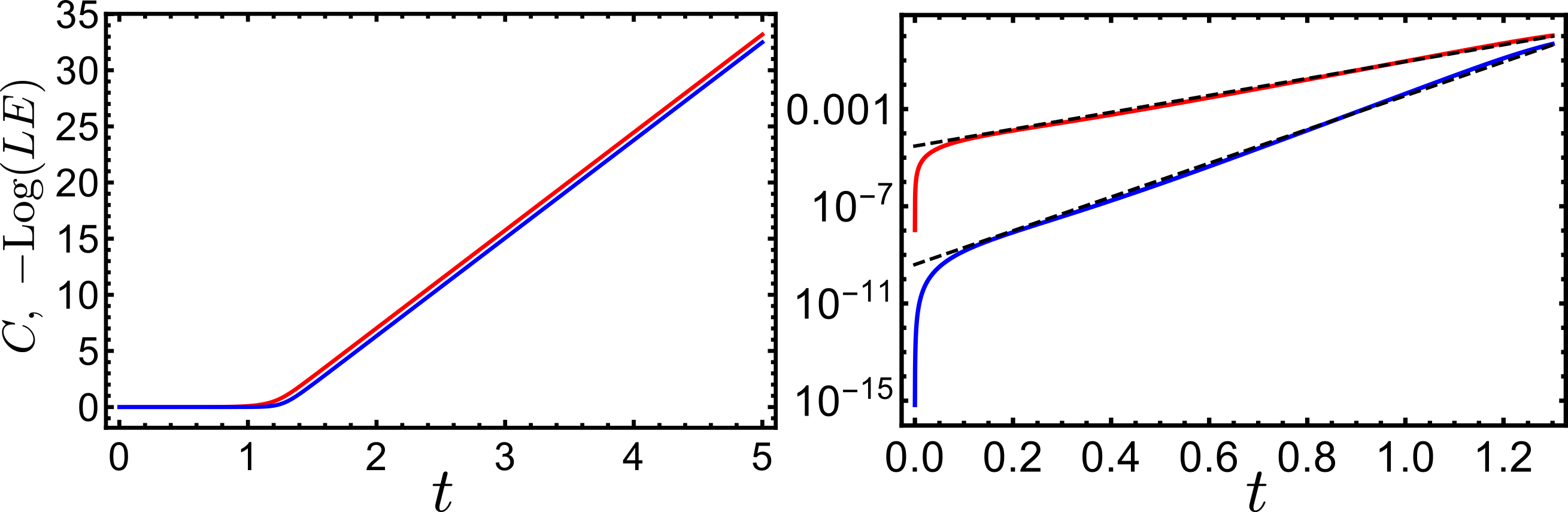}
\caption{Left: Time evolution of negative logarithm of LE and complexity for inverted oscillator ($m=1,\lambda=20,\delta\lambda=0.001$). Right: Logarithmic plot for early time behaviour of the negative logarithm of LE (blue) and complexity (red) and slope matching (dotted lines). }
\label{reg}
\end{figure*}

It is noteworthy that the construction procedure of this quantum circuit is conceptually similar to the LE, where one basically computes the overlap between these above mentioned states. Complexity simply offers us a different measure for the distance which is a more powerful measure for understanding various properties of quantum systems \cite{Bhattacharyya:2018bbv,ali2019time,Ali:2018aon,Bhattacharyya:2019kvj}.

As shown in Fig.~\ref{reg}, both of the time evolutions of the LE and complexity exhibit two regimes of growth, i.e., an intermediate regime where complexity grows linearly in time, while the LE decays as a pure exponential function (Fig.~\ref{reg}, left); an early regime (scrambling \cite{Maldacena:2015waa}) where complexity grows exponentially and the LE decays as a double exponential. The growth pattern of complexity and the LE are suggestive of the following relationship between LE and complexity, and their universal forms:

In the early scrambling regime, complexity and the LE have the form
\begin{equation}
\begin{aligned}
        \mathcal{C} =& ae^{\lambda t}\\
        \text{LE} =& ce^{-\epsilon e^{2\lambda t}}
\end{aligned}
\end{equation}

Figures~\ref{reg} confirms that during the scrambling stage, these two quantities are remarkably close. The same Lyapunov exponent $\lambda$, which is a system characteristic, can be extracted from both of these two quantities. The time scales of the early scrambling are also the same. Note also that the double exponential decay of the LE, when expanded to first order of $\epsilon$, reassembles the conjectured universal form of scrambling of the out-of-time ordered correlators of the form $1-\epsilon e^{\Lambda t}$. Hence, in the scrambling regime, we conjecture a universal relation between complexity and the LE:
\begin{equation}
\mathcal{C}^2 \sim -\log\ [\text{LE}].
\end{equation}

While in the intermediate regime, we have observed similar relations between the their growth rates, i.e., indicated by the evolution forms
\begin{equation}
\begin{aligned}
        \mathcal{C} =& a\,\Gamma t\\
        \text{LE} =& c\, e^{-a\Gamma t}
\end{aligned}
\end{equation}
Note that though they exhibit the same growth rate $\Gamma$, we expect this to be an artifact of the harmonic oscillator model. The exponential decay of the LE is standard feature and it is well-known \cite{goussev2012loschmidt,gorin2006dynamics} that its decay rate is a perturbation-dependent quantity, rather than a universal characteristic of the system.

\begin{figure*}[t] 
\centering
\includegraphics[width=0.9\textwidth]{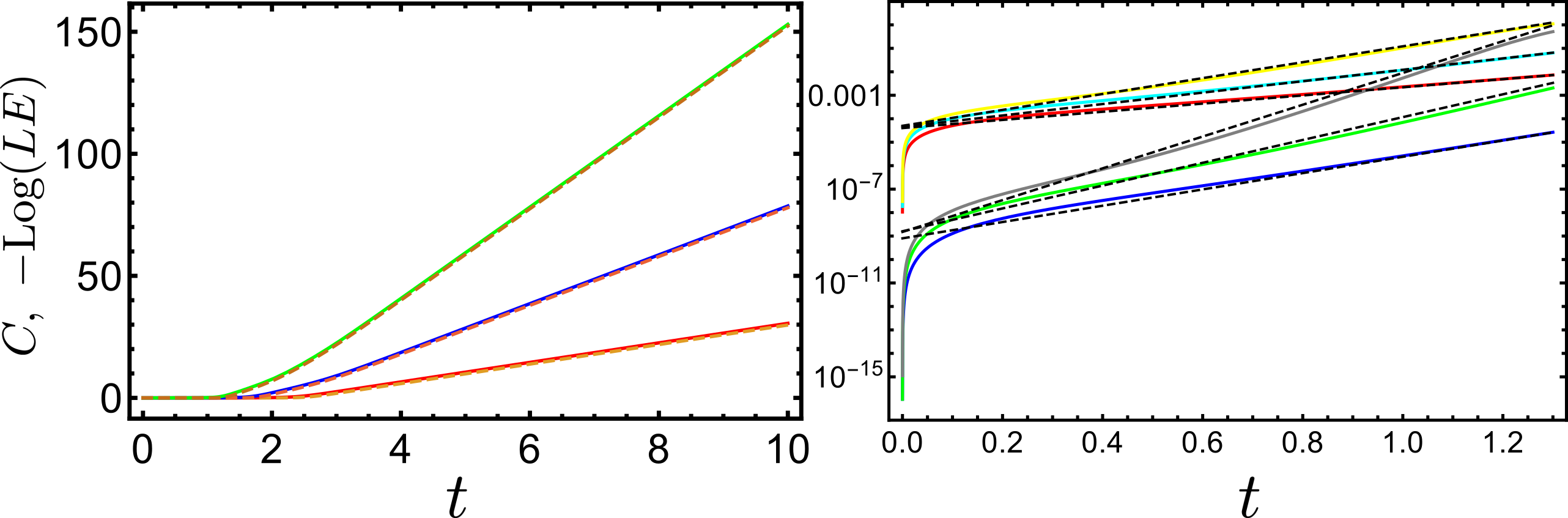}
\caption{Left: Time evolution of negative of logarithm of higher fold LE and complexity for Inverted Oscillator ($m=1,\lambda=5$ \ (Green), $10$\ (Purple), $20$\ (Red), $\delta\lambda=0.001$). The solid line is complexity and the dashed line is $ - \log$ [LE]. \ Right: Logarithmic plot for early time behaviour of higher LE (blue, green, gray) and complexity (red, cyan, yellow) and slope matching (black dotted lines).}
\label{reg2}
\end{figure*}

We can easily generalize this particular construction of quantum circuit to relate it with $2(k-1)$-fold LEs. The trick is to insert a pair of evolutions (forward and followed by a backward) for each fold of the echo. For example, for the 4-fold LE the quantum circuit we need to construct has the following form for the target state
\begin{equation}
  | \psi_4\rangle = \underbrace{e^{i (H'+\delta)  t}e^{-iH' t}}_{\text{2nd pair of evolutions}} \ \underbrace{e^{i (H+\delta)  t}e^{-iH t}}_{\text{1st pair of evolutions}} 
 |\psi_0\rangle .  
\end{equation}
In Fig.~\ref{reg2} we show a few of the higher fold-LE ($-\log$ \ [LE], to be precise) and the corresponding generalization of complexities. For each pair we see a clear match between complexity and the $- \log$ \ [LE] during the linear portion. The right panel of Fig.~\ref{reg2} displays the early time behaviour of these two quantities, which is similar to findings for the single fold case. We would like to stress that we do not have a concrete algebraic proof to establish the relationship at this point, therefore, it is just a conjecture and a concrete prove; we leave it for a future work.

Note that the sub-system LE that we have used in the previous sections can be quite close to full system LE, when the sub-system associated with the LE is much larger that the other one. We will conclude this section by making the assertion that these three diagnostics of chaos--averaged OTOC, LE and a particular type of complexity--are not only carrying similar information about the underlying quantum system, but also have some direct connection with each other.  

\section{Discussion}

In this paper we have extended the proof that the averaged (Haar average over unitaries) OTOC is the same as the LE (for a sub-system) as in \cite{zurek} to higher point averaged OTOC and LE for finite dimensional system. Moreover, we have also generalized the proof for Haar average to infinite dimensional case. We have shown that the OTOC-LE relation holds in other averaging scenarios as well, e.g., the Heisenberg group average, as long as the given group is a unitary 1-design. We argue that if the sub-system for this LE is much larger than the other sub-system, this LE would be essentially the same as the LE of the full system.  

Furthermore, for an explicit example in the Heisenberg group we showed graphically that LE for the full system and complexity for some special type of quantum circuit is the same. Finally, we have extended this result for multi-fold LE and corresponding extensions of the complexity. These different results suggest that these three diagnostics of a chaotic quantum system, namely averaged OTOC, LE and complexity are secretly the same. However, we do not have a concrete proof at this point. Tying complexity as an alternative probe to OTOC or LE also provides a geometric meaning to the chaotic behaviour of a quantum system.

To give a proof-of-principle argument for the similarity between complexity and LE, we have used the inverted oscillator as a toy model. This is, however, a rather special example and not a realistic chaotic system. Also, we used graphical techniques to establish our result. To claim that our particular complexity and LE (and hence averaged OTOC) are basically the same probe for understanding a quantum chaos will require a rigorous algebraic proof by using more ‘realistic’ systems like the maximally chaotic SYK model and its many variants (see, for example, \cite {Maldacena:2016hyu, Fu:2016vas,Kitaev:2017awl} and references therein).

Another possible extension of our work is to explore sub-system complexity in a system with N-inverted oscillators. This would help us make the connection between these quantities more rigorously. 

\begin{acknowledgements}
The authors would like to thank Aritra Banerjee, Jordan Colter, Jorge Kurchan and  Dan Roberts for useful discussions and email exchanges. W.A.C would like to thank the Institute for Quantum Information and Matter (IQIM), Caltech, for the ongoing stimulating environment from which the author has been significantly benefited.  A.B. is supported by  Start Up Research Grant (SRG/2020/001380) by Department of Science \& Technology Science and Engineering Research Board (India). W.A.C gratefully acknowledges the support of the Natural Sciences and Engineering Research Council of Canada (NSERC). B. Y. acknowledges support from the U.S. Department of Energy, Office of Science, Basic Energy Sciences, Materials Sciences and Engineering Division, Condensed Matter Theory Program, and partial support from the Center for Nonlinear Studies.
\end{acknowledgements}

\section*{Author Contributions}
All authors contributed equally to this paper. 

\begin{appendix}
\section{Mathematical Tools} 
Here we define the mathematical tools that will be handy in the derivations to be performed. Those tools include
\begin{enumerate}
    \item In Section II of the main text, a formula for the Haar average of a given trace-class operator $O$ has been discussed. At finite dimension
    \begin{equation}
        \int\limits_{Haar} dU(U^\dag)OU=\frac{1}{d}\, Tr(O){\rm I},
    \end{equation}
    where $I$ is the identity operator. The Hilbert space dimension $d$ appears here because of the Haar measure is normalized by convention, $\int dU =1$.
    At infinite dimension, a similar relation holds as well:
    \begin{equation}
        \int\limits_{Haar} dU(U^\dag)OU= {\rm Tr}(O)\, {\rm I},
    \end{equation}
    Here we present the formula for the Haar average of unitary operators restricted to a subsystem, which has been derived in Ref. \cite{zurek}. We only consider the finite dimensional case. For infinite dimensions similar relations can be treated in the same manner.
    \begin{equation}\begin{split}
   \int\limits_{Haar} dU_A(U_A^{\dagger}\otimes I_B O_{AB} U_A \otimes I_B)& = \int dU_A (U_A^{\dagger}\otimes I_B)(\sum\limits_{i} O_i^A \otimes O_i^B)(U_A\otimes I_B)\\
        & = \sum\limits_i \int dU_A (U_A^{\dagger} O^A_i U_A\otimes O_i^B)\\
  & = \frac{1}{d_A}\sum\limits_i {\rm Tr}(O_i^A)I_A\otimes O_i^B.
  \end{split}
  \end{equation}
  We have used  $$(A\otimes B)(C\otimes D) = (AC)\otimes (BD)$$ to get the first equality.  Further we used the following, 
  $$ k(A\otimes B)=A\otimes(kB)=kA\otimes B,$$ with $k$ being scalar we finally obtain that
  \begin{equation}\begin{split}
   \int\limits_{Haar} dU_A(U_A^{\dagger}\otimes I_B O_{AB} U_A \otimes I_B)&= \frac{1}{d_A}\sum\limits_i I_A \otimes {\rm Tr}(O_i^A)O_i^B \\
    & = \frac{1}{d_A} I_A\otimes {\rm Tr}_A O_{AB}.
       \end{split}
  \end{equation}
        
    \item Reduced dynamics for local operators \cite{zurek}: \\\\
    Given a total system Hamiltonian
    \begin{equation}
        H = H_A \otimes I_B + I_A \otimes H_B + H_I,
    \end{equation}
    where $A$ denotes a small local subsystem $S_A$. $B$ denotes the compliment of $S_A$ to the total system, which is much larger compared to the local system $S_A$. We are interested in strongly coupled systems, where the energy scales admits a hierarchy $\bar{H}_A\ll\bar{H}_I\ll\bar{H}_B$. For instance, in a $N$-particle system with all-to-all two-body interactions, when the subsystem $S_A$ refers to a single particle, the energy scales of $S_A$, $S_B$, and the coupling between them, are on the order of $1$, $N^2$ and $N$, respectively. The interaction can be decomposed as
    \begin{equation}
        H_I = \lambda \sum_{i=1}^{d_A^2} V_A^i\otimes V_B^i.
    \end{equation}
    Here we are free to chose the operators $\{V_A^i\}$ Hermitian and orthnormal, with respect to the Hilbert-Schmidt inner product, i.e.,
    \begin{equation}
        {\rm Tr}(V_A^iV_A^j)=d_A\delta_{i,j},
    \end{equation}
    where $d_A$ is the dimension of the Hilbert space of $S_A$. The operators $V_B^i$ on $S_B$ are also Hermitian, but their (Hilbert-Schmidt) norms are fixed as equal to the norms of $H_B$. Thus, the parameter $\lambda$ qualifies the relative strength of the coupling compared to $H_B$.
    
    We are interested in the reduced dynamics of an operator $B$ on the subsystem $S_B$, after the trace-out procedure, namely,
    \begin{equation}
    B(t) = {\rm Tr}_A \left(e^{i H t} I_A\otimes B e^{-i H t}\right). 
    \end{equation}
    This can be thought of as a decoherence process, i.e., the total system is prepared in an initial product state $I_A\otimes B$, where the subsystem $S_B$ has a ``density matrix'' $B$, and the subsystem $S_A$, up-to normalization, is in a thermal state with infinite temperature. The ``quantum state'' $B$ will become ``mixed'' with time evolution due to the presence of the couplings to subsystem $S_A$. When $\lambda\ll 1$, the above evolution of $B(t)$ can be expanded to the second order of $\lambda$. This corresponds to the Born-Markov approximation, which leads the effective master equation for $B(t)$ to a Lindblad form. It is known that in this case the effective master equation can be simulated with the evolution of $B$ under $H_B$ without coupling to other systems, but subjects to a stochastic field
    \begin{equation}
        \lambda \mathcal{F}(t)=\lambda \sum_i l_i(t) V_B^i,
    \end{equation}
    with the correlations given by
    \begin{equation}
    \begin{aligned}
        &\ll  l_i(t)\ l_j(t-\tau) \gg \\
        = &{\rm Tr}(\frac{I_A}{d_A}V_A^i e^{i H_A \tau}V^j_A e^{-i H_A \tau})\\
        \approx &\delta_{i,j}.
    \end{aligned}
    \end{equation}
    The approximation in the last step is due to the large energy hierarchy: the time scale of the dynamics of the subsystem $S_A$ is much larger than that of $B(t)$ under consideration. Alternatively, this can be thought of as taking the zeroth order the $H_A$. 
    As a consequence, the noise field $l_i(t)$ can be taken as random constant valued, $\pm 1$, at equal probability. The reduced dynamics of the $B$ operator is then given by
    \begin{equation}
        B(t)=d_A \ll e^{-i (H_B + \lambda\mathcal{F})t } B e^{i (H_B + \lambda\mathcal{F})t } \gg,
    \end{equation}
    averaged over the stochastic field. Note that the pre-factor $d_A$ appears from the normalization of $I_A$. As the noise field are random $\pm 1$, each realization of the stochastic field $\mathcal{F}$ in the above solution of $B(t)$ always appears as random combination of $V^i_B$'s. Suppose that are totally $N$ realizations, the noisy evolution of $B(t)$ is then 
\begin{equation}
        B(t) \approx d_A\times \frac{1}{N}\sum_{i,j=1}^{N} e^{-i (H_B + \lambda\mathcal{F}_k) t } B e^{i (H_B + \lambda\mathcal{F}_k)t }.
\end{equation}
    
    \end{enumerate}
\end{appendix}

\bibliography{references} 
\end{document}